\documentclass[aps,prb,reprint]{revtex4-2}
\usepackage{amsmath,amssymb,graphicx,numprint,newtxtext,newtxmath,dcolumn}
\usepackage[dvipsnames,svgnames,table]{xcolor}
\usepackage[colorlinks=true, urlcolor=blue, linkcolor=blue, citecolor=blue, pdftex]{hyperref}
\usepackage{orcidlink}
\npdecimalsign{.}

\def\<{\langle}
\def\>{\rangle}

\newcolumntype{b}{D{.}{.}{3}}
\newcolumntype{e}{D{.}{.}{4}}
\newcolumntype{f}{D{.}{.}{8}}
\newcolumntype{a}{D{.}{.}{6}}

\newcommand{\PRL}{Phys.~Rev.~Lett.}

\newcommand{\PRE}{Phys.~Rev. E}
\newcommand{\PRX}{Phys.~Rev. X}

\newcommand{\NPB}{Nucl.~Phys.~B}

\begin{document}

\title{Universal finite-size scaling in the extraordinary-log boundary phase \\of three-dimensional \texorpdfstring{$O(N)$}{O(N)} model}
\author{\firstname{Francesco} \surname{Parisen Toldin}\,\orcidlink{0000-0002-1884-9067}}
\email{parisentoldin@physik.rwth-aachen.de}
\affiliation{Institute for Theoretical Solid State Physics, RWTH Aachen University, Otto-Blumenthal-Str. 26, 52074 Aachen, Germany}
\affiliation{JARA-FIT and JARA-CSD, 52056 Aachen, Germany}
\author{\firstname{Abijith} \surname{Krishnan}}
\email{abijithk@mit.edu}
\author{\firstname{Max~A.} \surname{Metlitski}}
\email{mmetlits@mit.edu}
\affiliation{Department of Physics, Massachusetts Institute of Technology, Cambridge, Massachusetts 02139, USA}

\begin{abstract}
Recent advances in boundary critical phenomena have led to the discovery of a new surface universality class in the three-dimensional $O(N)$ model. The newly found ``extraordinary-log" phase can be realized on a two-dimensional surface for $N< N_c$, with $N_c>3$, and on a plane defect embedded into a three-dimensional system, for any $N$.
One of the key features of the extraordinary-log phase is the presence of logarithmic violations of standard finite-size scaling.
In this work we study finite-size scaling in the extraordinary-log universality class by means of Monte Carlo simulations of an improved lattice model.
We simulate the model with open boundary conditions, realizing the extraordinary-log phase on the surface for $N=2,3$, as well as with fully periodic boundary conditions and in the presence of a plane defect for $N=2,3,4$.
In line with theory predictions, renormalization-group invariant observables studied here exhibit a logarithmic dependence on the size of the system. We numerically access not only the leading term in the $\beta$-function governing these logarithmic violations, but also the subleading term, which controls the evolution of the boundary phase diagram as a function of $N$.

\end{abstract}
\maketitle

\section{Introduction}
\label{sec:intro}
The last several years have seen advances in our understanding of boundaries and defects in critical systems. There are several factors contributing to an interest in this topic, including questions about the existence of topologically protected boundary states in gapless quantum systems \cite{GV-12,BQ-14,BMF-15,CCBCN-15, SPV-17, PSV-18,TVV-21,Verresen-20, VTJP-21, YHSXDZ-22, RuiPotterCoho, RuiPotterHolog}, appearance of conical defects in the study of entanglement properties \cite{Calabrese_04}, mapping of questions about open quantum systems to more conventional statistical mechanics problems with defects \cite{ehudmeasure,CenkeMeasure}, and the use of extended operators to characterize phases of matter \cite{cordovaSM}. 

One unexpected recent development has been the theoretical prediction of a hitherto overlooked boundary phase, the so-called extraordinary-log phase \cite{Metlitski-20}. This phase appears on the two-dimensional boundary of the classical three-dimensional (3D) $O(N)$ model, when the bulk is tuned to the critical point,  the surface coupling is sufficiently large and the number of order parameter components $N$ satisfies $2 \leq N < N_c$. Here $N_c$ is a critical number of components estimated as $N_c \sim 5$ \cite{PKMGM-21}.
This is a surprising result, given that for $N>2$ the topology of the surface-bulk phase diagram does not require the existence of an extraordinary phase on the surface.
The extraordinary-log phase is characterized by a surface order parameter correlation function that decays extremely slowly as $\langle \vec{n}(x) \cdot \vec{n}(0)\rangle \sim 1/(\log x)^{q}$, where $q$ is an $N$-dependent universal number. If one describes the surface by a non-linear $\sigma$-model 
\begin{equation}
    S_{\text{bdy}} = \frac{1}{2g} \int d^2x \left(\partial_\mu\vec{n}\right)^2,\qquad \vec{n}^2=1,
\label{nls}
\end{equation}
then for $2 \leq N < N_c$ the interaction of the surface with the critical 3D bulk effectively reverses the RG flow of the coupling $g$ in Eq.~(\ref{nls}) compared to a purely 2D system and gives rise to a logarithmic flow to a stable fixed point at $g = 0$, which describes the extraordinary-log universality class (UC).

The same kind of behavior occurs in the 3D $O(N)$ model with a plane defect, except here the extraordinary-log phase exists for any $N \ge 2$ \cite{KM-23}. In both the boundary and plane defect case, as the surface coupling is reduced the surface goes through a phase transition from the extraordinary-log to the ordinary UC \cite{Diehl-86}, characterized by conventional boundary conformal field theory behavior. The transition point between the ordinary and extraordinary-log phases is dubbed the special UC.\cite{Diehl-86}\footnote{In the case of the plane defect, the special UC corresponds to the surface coupling equal to the bulk coupling, i.e a trivial defect in the 3D bulk.} 

\begin{figure*}
    \centering
    \includegraphics[width=0.45\linewidth]{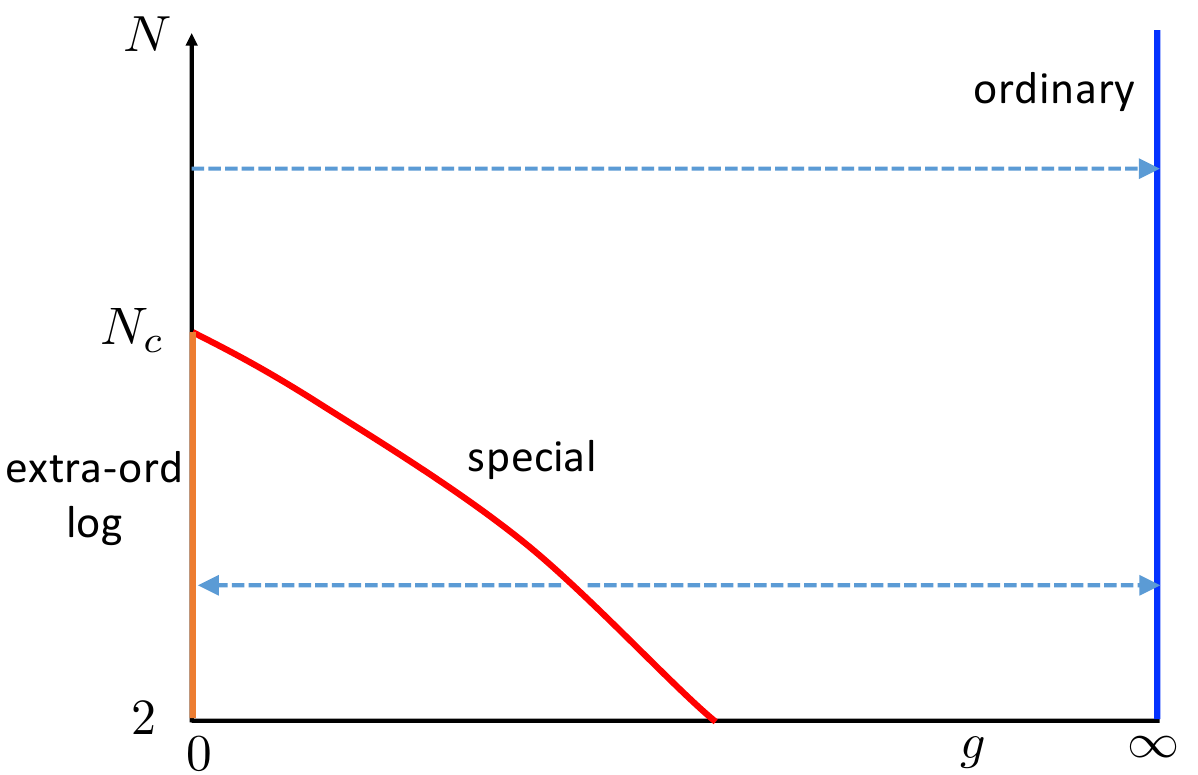} \includegraphics[width=0.45\linewidth]{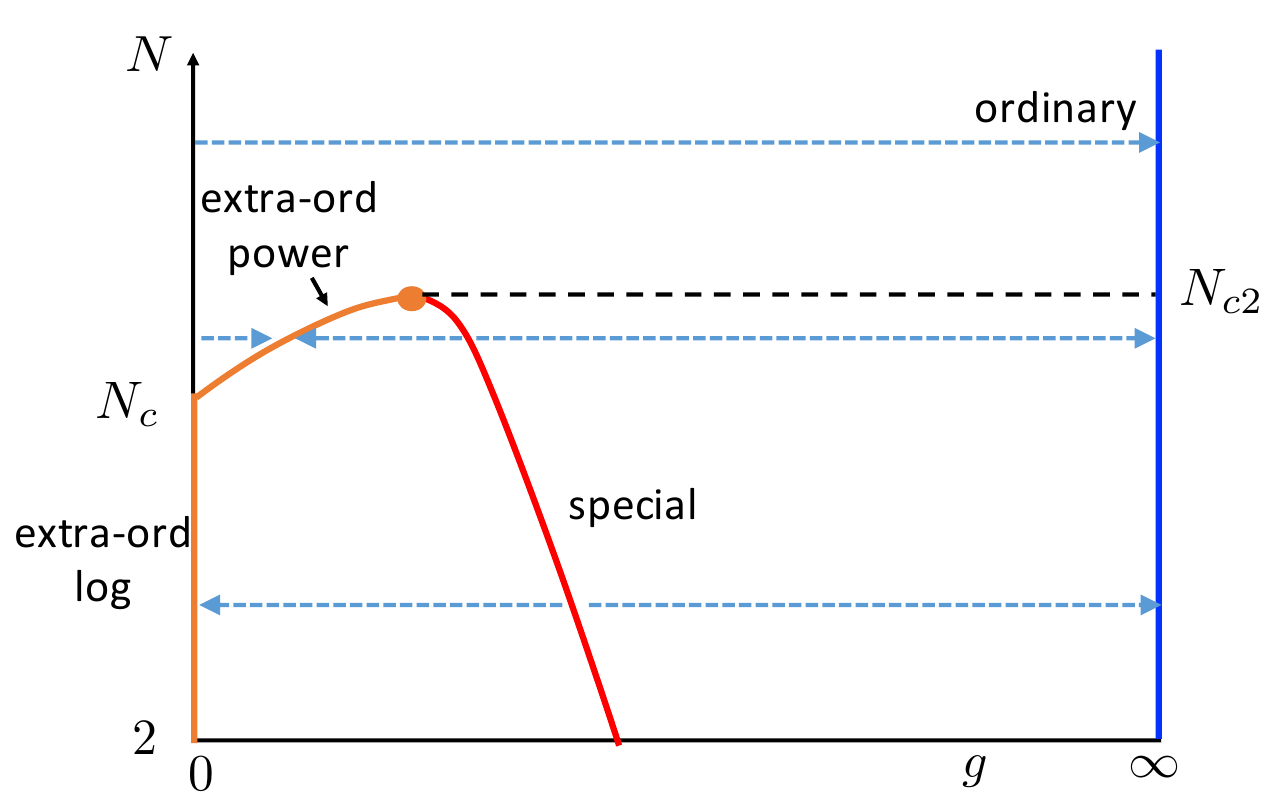}
    \caption{Surface RG flow of the 3D $O(N)$ model in the regime of large surface exchange coupling (small $g$). The left figure corresponds to the parameter $b$ in the $\beta$-function (\ref{beta}) satisfying $b(N_c) < 0$ and the right figure to $b(N_c) > 0$. In the right figure the extra-ordinary fixed point at $N_c < N < N_{c2}$  is labeled as ``extra-ordinary power" since it is characterized by boundary order parameter correlations falling off in a conventional power law manner.  Taken from  Ref.~\cite{Metlitski-20}. }
    \label{fig:RG}
\end{figure*}

The analytical predictions of the existence of the extraordinary-log boundary and defect UCs in the 3D $O(N)$ model made in Refs.~\cite{Metlitski-20,KM-23} were verified by Monte Carlo (MC) simulations \cite{PT-20, PTM-21, HDL-21, SHL-23}. In particular, Refs.~\cite{PT-20, PTM-21} have found that the extraordinary-log boundary UC exists in the $O(3)$ model, placing a bound $N_c > 3$.

One question that has not been fully resolved to date is the evolution of the boundary phase diagram in the $O(N)$ model as $N$ crosses $N_c$. Reference \cite{Metlitski-20} noted that this evolution is controlled by the sign of the first subleading term in the $\beta$-function of the effective surface coupling $g$ in Eq.~(\ref{nls}):
\begin{equation} \beta(g) = -\frac{dg}{d\ell} = \alpha(N) g^2 + b(N) g^3 + O(g^4). \label{beta} \end{equation}
Here $\ell$ is the RG scale. The coefficient of the leading term $\alpha$ in Eq.~(\ref{beta}) switches sign from positive to negative as $N$ increases from below $N_c$ to above $N_c$, so that the extraordinary-log fixed point $g = 0$ becomes unstable for $N > N_c$. The sign of the subleading term $b$ controls the surface physics in the vicinity of $N = N_c$. In particular, if $b(N_c) < 0$ then for $N$ slightly below $N_c$ there is a perturbatively accessible IR unstable fixed point $g_* \approx \frac{\alpha}{|b|}$  that is expected to correspond to the special UC. As $N$ approaches $N_c$ from below, the special fixed point approaches the extraordinary-log fixed point $g=0$, annihilating with it at $N_c$, so that only the ordinary fixed point is stable for $N > N_c$; see Fig.~\ref{fig:RG}, left. On the other hand, if the coefficient $b(N_c) > 0$, the evolution of the surface phase diagram for $N \ge N_c$ is more complicated:  the extraordinary fixed point moves to a finite value of the coupling $g_* \approx \frac{|\alpha|}{b}$ for $N$ just above $N_c$ and presumably annihilates with the special fixed point at a higher critical value $N_{c2} > N_c$, see Fig.~\ref{fig:RG}, right. While it was shown in Ref.~\cite{KM-23} that for $N \to \infty$, $b < 0$, no information about the sign of $b$ at $N = N_c$ has been available to date.

The  goal of the present paper is to perform a detailed finite-size scaling (FSS) analysis of the extraordinary-log UC. We consider the $O(N)$ model in the open boundary geometry for $N = 2,3$ and with a plane defect for $N = 2,3,4$.  We use MC simulations to study the behavior of several renormalization-group (RG) invariants, such as the system stiffness, correlation length and Binder ratio, and find good agreement with analytical predictions that we present. 
Our analysis allows us to extract the leading coefficient $\alpha$ in the $\beta$-function (\ref{beta}) more precisely than in previous direct studies of the extraordinary-log phase. Note that  $\alpha$ was theoretically predicted  to be determined by ratios of certain bulk to boundary operator product expansion (OPE) coefficients of the normal (fixed spin) boundary UC.\cite{Metlitski-20, PTM-21, PKMGM-21, KM-23}\footnote{The normal UC is accessed by turning on a perturbation that explicitly breaks the $O(N)$ symmetry to $O(N-1)$ on the boundary.} These OPE coefficients were extracted using MC simulations in Ref.~\cite{PTM-21} and give a value of $\alpha$ that is in excellent agreement with the results  we present here. Moreover, our analysis allows us to estimate the value of the subleading coefficient $b(N)$ in the $\beta$-function (\ref{beta}). In particular, for the model with a boundary our results indicate $b(N = 3) < 0$. As $b(N =\infty)$ is analytically known to be negative as well, one expects that $b(N)$ remains negative at $N = N_c$,  which  corresponds to the scenario for the evolution of the boundary phase diagram in Fig.~\ref{fig:RG}, left. We also assess whether for open boundary conditions (BCs) the critical properties of the special transition found with MC simulations in Refs.~\cite{PT-20, HDL-21, Deng-06} are consistent with a perturbative expansion that is expected to become accurate as $N \to N^-_c$. We find that, indeed, for $N = 3, 4$ the special transition appears to lie in a perturbatively accessible regime.

This paper is organized as follows. In Sec.~\ref{sec:fss} we present theoretical predictions for FSS of the stiffness, correlation length and Binder ratio in the extraordinary-log phase. Section \ref{sec:results} defines the ``improved" lattice model we study with MC and presents the FSS simulation results. Implications of these results for the evolution of the boundary phase diagram of the $O(N)$ model with $N$ are discussed in Sec.~\ref{sec:outlook}. Supplementary details are relegated to the Appendixes. 
In Appendix \ref{app:U4} we analytically compute the FSS behavior of the RG-invariant Binder ratio in the extraordinary-log phase.
In Appendix \ref{app:beta_s1.5} we present some additional results of our MC FSS analysis.
In Appendix \ref{app:beta_O32d}, as a test, we apply the method used in the main text to extract the surface $\beta$-function to the O($3$) model in two dimensions, where the leading terms in the $\beta$-function are analytically known.

\section{Finite-Size Scaling}
\label{sec:fss}

\begin{figure}[t]
\centering
 \begin{tikzpicture}
  \draw[dashed, fill = violet!40] (2,0) -- (2.5, 0.7) -- (0.5, 0.7) -- (0,0) -- (2,0);
 \filldraw[gray!30, opacity = 0.5](0,0) -- (2,0) -- (2, 2) -- (0,2) -- (0,0);
 \filldraw[gray!50, opacity = 0.5](2,0) -- (2.5,0.7) -- (2.5,2.7) -- (2,2) -- (2,0);
\draw[dashed](0.5,0.7) -- (0.5,2.7);
\draw[dashed](0,0) -- (0,2);
\draw[dashed](2.5,2.7) -- (2.5,0.7);
\draw[dashed](2,0) -- (2,2);
 \draw[dashed, fill = violet!40, opacity=0.7] (2,2) -- (2.5, 2.7) -- (0.5, 2.7) -- (0,2) -- (2,2);

\node at (1.25, 1.35) {$\beta$}; 
\node at (1.25, 2.35) {$\beta_s$}; 
\node at (1.25, 0.35) {$\beta_s$}; 

\node at (1,-0.5) {Open BCs};
 \end{tikzpicture}
  \hspace{0.1\textwidth} 
\begin{tikzpicture}
  \draw[dashed] (2,0) -- (2.5, 0.7) -- (0.5, 0.7) -- (0,0) -- (2,0);
 \draw[dashed, fill = orange!40] (2,1) -- (2.5, 1.7) -- (0.5, 1.7) -- (0,1) -- (2,1);
 \filldraw[gray!30, opacity = 0.5](0,0) -- (2,0) -- (2, 2) -- (0,2) -- (0,0);
 \filldraw[gray!50, opacity = 0.5](2,0) -- (2.5,0.7) -- (2.5,2.7) -- (2,2) -- (2,0);
 \draw[dashed](0.5,0.7) -- (0.5,2.7);
\draw[dashed](0,0) -- (0,2);
\draw[dashed](2.5,2.7) -- (2.5,0.7);
\draw[dashed](2,0) -- (2,2);
  \filldraw[gray!20, opacity = 0.5] (2,2) -- (2.5, 2.7) -- (0.5, 2.7) -- (0,2) -- (2,2);
   \draw[dashed] (2,2) -- (2.5, 2.7) -- (0.5, 2.7) -- (0,2) -- (2,2);
\node at (1, 0.5) {$\beta$}; 
\node at (1.25, 1.35) {$\beta_s$}; 
\node at (1,-0.5) {Plane defect};
 \end{tikzpicture}
\caption{The geometry of the two realizations of the extraordinary log UC of the 3D O$(N)$ model, open BCs (left) and plane defect (right). The gray surfaces have periodic BCs, the violet surfaces have open BCs. The bulk $O(N)$ coupling is $\beta$ and the surface $O(N)$ coupling is $\beta_s$ [see Eq.\ \eqref{model}].}
\label{fig:geometry}
\end{figure}
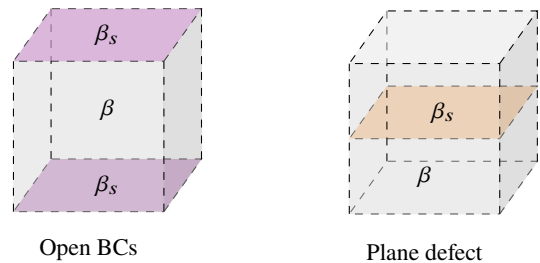

In this paper, we consider two 3D geometries that realize the extraordinary-log UC: a system with periodic BCs in two directions and open BCs for two confining surfaces that realize the extraordinary-log UC, and a fully periodic system with a plane defect on which the extraordinary-log UC is realized (see Fig.\ \ref{fig:geometry}). 
In both cases, we consider systems with equal size $L$ along the three directions.

Following Ref.~\cite{Metlitski-20}, we describe the boundary/defect physics with a nonlinear $\sigma$-model
for the $N-$component field $\vec{n}$ in Eq.~(\ref{nls}).
This $\sigma$-model is coupled to the 3D critical $O(N)$ model with a normal (fixed spin) boundary condition in a universal manner that restores the $O(N)$ symmetry.
The RG flow of $g$ in the vicinity of the $g=0$ fixed point is given by Eq.~(\ref{beta}), 
where the universal RG parameter $\alpha$ is determined by the coupling to the critical bulk, and can also be extracted from particular universal amplitudes in the normal UC \cite{Metlitski-20,PTM-21, PKMGM-21}.
The extraordinary-log phase is realized when $\alpha>0$.
In this case $g$ flows logarithmically to 0 as
\begin{equation}
    \frac{1}{g(\ell)} = \frac{1}{g(\ell_0)} + \alpha \ell + \ldots,
\label{gflow}
\end{equation}
where the leading suppressed term in (\ref{gflow}) originates from the $bg^3$ correction in (\ref{beta}) and scales as $\log \ell$ for $\ell \to \infty$.

We study the FSS behavior of several RG-invariants.
First, we compute the helicity modulus $\Upsilon$ by introducing a torsion of angle $\theta$ on a single lateral direction, where periodic BCs are applied, and a single plane in the $O(N)$ space.
Then, $\Upsilon$ is the response of the system to the torsion, and it is defined as \cite{FBJ-73}
\begin{equation}
  \Upsilon \equiv \frac{L}{S} \frac{\partial^2 F(\theta)}{\partial \theta^2}\Big|_{\theta=0},
  \label{helicitydef}
\end{equation}
where $S=L_\parallel^2=L^2$ is the area of the surfaces, and $F$ is the free energy in units of $k_BT$.
With such a definition, $L\Upsilon$ is a RG-invariant quantity.
When $g \to 0$, its FSS is dominated by the boundary contribution.
Ignoring fluctuations, each boundary contributes as $(2/N)g^{-1}$ \cite{MW-01}.
Since $L \Upsilon$ is an RG-protected quantity, we can RG improve it by simply using $g \to g(L)$.
Employing Eq.~(\ref{gflow}), we arrive at
\begin{alignat}{2}
      L \Upsilon  &\approx \frac{4}{Ng(L)}  \approx  L_0 \Upsilon(L_0) + \frac{4 \alpha}{N} \log L/L_0, &\quad \text{open BCs},
      \label{FSSLY_open}\\
      L \Upsilon  &\approx \frac{2}{Ng(L)}  \approx  L_0 \Upsilon(L_0) + \frac{2\alpha}{N} \log L/L_0, &\quad \text{plane defect},
      \label{FSSLY_plane}
\end{alignat}
where we have included the  $O(g^0)$  contribution into $L_0 \Upsilon(L_0)$. Here $L_0$ is a reference length scale.
We point out that there was an error in the expression for the stiffness in Ref.~\cite{Metlitski-20}--- the prefactor $2/N$ was missing.
We note that the critical bulk gives rise to an additional contribution to $L\Upsilon$ which is of $O(g^0)$, cumulating to the $L_0 \Upsilon(L_0)$ term and subleading with respect to the logarithmic divergence.

Another useful RG-invariant quantity is the
ratio $\xi/L$, where $\xi$ is the finite-size correlation length  on the boundary.
This quantity is defined as \cite{CP-98,PTHAH-14}
\begin{equation}
\left(\frac{\xi}{L}\right)^2 = \frac{1}{(2 L \sin(\pi /L))^2} \left(\frac{\tilde{C}(0)}{\tilde{C}(2\pi/L)} - 1\right),
\label{xildef}
\end{equation}
with 
\begin{equation}
\tilde{C}(\vec{p}) = \int d^2 x \, C(\vec{x}) e^{-i \vec{p} \cdot \vec{x}}
\end{equation}
and
$C(\vec{x})$ is the two-point function of the order parameter on the boundary;
for the nonlinear $\sigma-$model of Eq.~(\ref{nls}) $C(\vec{x})= \langle \vec{n}(\vec{x}) \cdot \vec{n}(\vec{0})\rangle$.
Following Ref.~\cite{Metlitski-20}, we express the vector field $\vec{n}$ as $\vec{n} = (\vec{\pi}, \sqrt{1-\vec{\pi}^2})$, where the $(N-1)$-component vector field $\vec{\pi}$ describes the fluctuations around the symmetry-broken fixed point $g=0$ and satisfies $\int d^2 x \, \vec{\pi}(\vec{x}) = 0$.
In doing this, we have assumed without loss of generality that at $g=0$ the order parameter points in the $N$ direction.
Using the action (\ref{nls}) we obtain, to first order in $g$
\begin{equation}
C(\vec{x}) = 1 + (N-1) (D_{\pi}(\vec{x}) - D_{\pi}(0))
\end{equation}
with
\begin{equation}
D_{\pi}(\vec{x})  = \frac{1}{L^2} \sum_{\vec{p} \neq 0} \frac{g}{\vec{p}^2} e^{i \vec{p} \cdot \vec{x}}\end{equation}
To zeroth order in $g$, we just have $C(\vec{x}) = 1$, i.e. $\tilde{C}(\vec{p}) = L^2 \delta_{\vec{p},0}$. Thus, while the leading contribution to $\tilde{C}(0)$
\begin{equation}
\tilde{C}(0) \approx L^2,
\label{C0}
\end{equation}
is $O(g^0)$, the leading contribution to $\tilde{C}(\vec{p})$ with $\vec{p} \neq 0$ is $O(g^1)$ 
\begin{equation}
\tilde{C}(\vec{p}) \approx (N-1) \frac{g}{p^2},
\end{equation}
so that
\begin{equation}
\frac{\tilde{C}(0)}{\tilde{C}(p)} \approx \frac{1}{(N-1)g} (pL)^2.
\label{Cratio}
\end{equation}
Inserting Eq.~(\ref{Cratio}) in Eq.~(\ref{xildef}) we obtain
\begin{equation}
\left(\frac{\xi}{L}\right)^2 \approx \frac{1}{(N-1) g},
\label{xiLunimp}
\end{equation}
where we have dropped 
the term $-1/(2 L \sin(\pi/L))^2 \approx -1/4\pi^2$ in Eq.~(\ref{xildef}), since in Eq.~(\ref{C0}) we are already neglecting other terms of the same magnitude.
$\xi/L$ is an RG protected quantity, so we may RG improve it by the substitution $g \to g(L)$:
\begin{equation}
\begin{split}
   \left(\frac{\xi}{L}\right)^2 &\approx \frac{1}{(N-1) g(L)} = \frac{1}{(N-1) g(L_0)} + \frac{\alpha}{N-1} \log L/L_0 \\
   &= \left(\frac{\xi(L_0)}{L_0}\right)^2  + \frac{\alpha}{N-1} \ln (L/L_0).
\end{split}
\label{FSSxil}
\end{equation}
We note that $\alpha$ also determines the asymptotic behavior of $C(x)$ in real space \cite{Metlitski-20}
\begin{equation}
C(x) = \frac{1}{(\log x)^{q}}, \qquad q = \frac{N-1}{2 \pi \alpha}.
\end{equation}
Thus, the coefficients of the $\log L$ terms in $L \Upsilon$ [Eqs.~(\ref{FSSLY_open}) and (\ref{FSSLY_plane})] and $(\xi/L)^2$ [Eq.~(\ref{FSSxil})] are both related to $\alpha$.

We finally study another commonly employed RG-invariant quantity, the Binder ratio $U_4$, defined on a boundary as
\begin{equation}
U_4 = \frac{M_4}{M^2_2},
\label{U4def}
\end{equation}
where
\begin{equation}
M_{2p} = \left\langle \left(\int d^2 x d^2 y \, \vec{n}(x)\cdot \vec{n}(y)\right)^p  \right\rangle
\end{equation}
A perturbative calculation in the non-linear $\sigma$-model similar to the one for $\xi/L$ above gives (see appendix \ref{app:U4}): 
\begin{equation}
U_4 - 1 \approx \frac{2 (N-1) {\cal S} g^2}{(2\pi)^4}, \quad\quad {\cal S} \approx 6.02681.
\label{U4g0}
\end{equation}
Since $U_4$ is an RG protected quantity, we may improve it as
\begin{equation}
\begin{split}
U_4 - 1 &\approx \frac{2 (N-1) {\cal S} g(L)^2}{(2\pi)^4}\\
&\approx \frac{2 (N-1) {\cal S}}{(2\pi)^4} \frac{g(L_0)^2}{(1 + \alpha g(L_0) \log(L/L_0))^2}, \label{FSSU40}
\end{split}
\end{equation}
i.e., for asymptotically large $L$, 
\begin{equation}
U_4 - 1 \approx  \frac{2 (N-1) {\cal S}}{(2 \pi)^4 \alpha^2 \log^2(L)}.
\label{FSSU4}
\end{equation}

We note that for the quantities ${\Upsilon} L$, $\xi/L$ and $U_4 - 1$ that we have computed so far, to leading order in $g$ the coupling of the nonlinear $\sigma$-model to the bulk can be ignored. This coupling enters only implicitly, when we RG improve our results.

We also   compute combinations of observables where the $g$ dependence drops out. For instance,
\begin{equation}
     (U_4 - 1)\left(\xi/L\right)^4 \underset{L\rightarrow\infty}{\approx} \frac{{\cal S}}{8 \pi^4 (N-1)},
    \label{FSSxilU4}
\end{equation}
and
\begin{alignat}{2}
    (L \Upsilon) \left(\xi/L\right)^{-2} &\underset{L\rightarrow\infty}{\approx} 4 (N-1) / N, \quad &\text{open BCs},
    \label{FSSLYxil_open}\\
    (L \Upsilon) \left(\xi/L\right)^{-2} &\underset{L\rightarrow\infty}{\approx} 2 (N-1) / N, &\text{plane defect}.
    \label{FSSLYxil_plane}    
\end{alignat}

\section{Results}
\label{sec:results}
\subsection{Model}
\label{sec:results:model}
To realize the extraordinary-log UC, we simulate the classical $\phi^4$ lattice model.
Its reduced Hamiltonian ${\cal H}$, defined such that the Gibbs weight is $\exp(-\cal H)$, is
\begin{equation}
    {\cal H} = -\beta\sum_{\< i\ j\>}\vec{\phi}_i\cdot\vec{\phi}_j
    -\beta_s\sum_{\< i\ j\> \in \partial}\vec{\phi}_i\cdot\vec{\phi}_j
    +\sum_i[\vec{\phi}_i^{\,2}+\lambda(\vec{\phi}_i^{\,2}-1)^2],
  \label{model}
\end{equation}
where $\vec{\phi}_x$ is an $N-$component real field on the lattice site $x$.
In Eq.~(\ref{model}), the second sum extends over the nearest-neighbor pairs on the boundary, i.e., on the two surfaces in the case of open BCs, or on the plane defect, whereas the first sum extends over all other nearest-neighbor sites; the third term is summed over all lattice sites.
In the thermodynamic limit, the Hamiltonian (\ref{model}) shows a critical line  $(\lambda, \beta_c(\lambda))$ in the 3D O($N$) UC, along which it is possible to find, for $N\le 4$, a single critical point $(\lambda^*, \beta_c(\lambda^*))$ where the model is improved, i.e., the leading scaling correction vanishes.
In this work, as we have done in recent MC studies of boundary critical phenomena \cite{PT-20,PTM-21,PT-23}, we study the model for $N=2,3,4$, fixing $\beta$ and $\lambda$ to the improved critical point;
the corresponding parameters used in the simulations are: $\lambda=2.15$, $\beta=\numprint{0.50874988}$ \cite{PT-21} for $N=2$, $\lambda=5.2$, $\beta=\numprint{0.68798521}$ \cite{Hasenbusch-20} for $N=3$, and $\lambda=18.5$, $\beta=\numprint{0.91787555}$ \cite{Hasenbusch-21} for $N=4$.

The coupling constant $\beta_s$ in Eq.~(\ref{model}) controls the strength of the interaction on the boundary, and potentially
 induces the extraordinary-log UC.
More precisely, in the case of open BCs and for $N\le N_c$ (with $N_c \approx 5$ \cite{PKMGM-21}), the extraordinary-log UC is obtained when $\beta_s>\beta_{s,c}$, with $\beta_{s,c}$ the critical coupling at the special transition \cite{Metlitski-20}.
For plane defect geometry, the extraordinary-log UC is realized when $\beta_s>\beta_c$ for any $N$ \cite{KM-23}; in this case the special UC corresponds to a translationally invariant system, and the critical exponents on the plane defect coincide with the bulk ones.
For both BCs studied here, it is crucial to consider values of the boundary interaction strength $\beta_s$ large enough,  otherwise the proximity to the special transition induces undesirable large crossover effects \cite{PT-20}.

\begin{table}[t]
    \centering
    \caption{Value of the universal RG parameter $\alpha$ governing the RG flow in the vicinity of the extraordinary-log UC. The reported values are taken from Ref.~\cite{PTM-21} ($N=2,3$) and Ref.~\cite{O4new} ($N=4$).
    }
    \begin{ruledtabular}
    \begin{tabular}{caa}
    $N$ & \multicolumn{1}{c}{open BCs} & \multicolumn{1}{l}{plane defect} \\
    \hline
         $2$ & 0.300(5) & 0.600(10) \\
         $3$ & 0.190(4) & 0.540(8) \\
         $4$ & 0.097(3) & 0.512(6) \\
    \end{tabular}
    \end{ruledtabular}
    \label{tab:alpha}
\end{table}
As detailed in Sec.~\ref{sec:fss}, FSS in the extraordinary-log phase is controlled by the universal RG parameter $\alpha$, which is determined by some amplitudes in the normal UC.
In Table \ref{tab:alpha} we report estimates of $\alpha$ for both geometries considered here, and $N=2,3,4$, as obtained in Refs.~\cite{PTM-21,O4new}.
Since $\alpha$ enters as a prefactor to the logarithmic terms [Eqs.~ (\ref{FSSLY_open}), (\ref{FSSLY_plane}) (\ref{FSSxil}) and (\ref{FSSU40})], clearly a small value of $\alpha$ renders it difficult to numerically check the FSS behavior discussed in Sec.~\ref{sec:fss}.
Therefore, based on the values reported in Table \ref{tab:alpha}, we choose to study here the extraordinary-log UC in open BCs for $N=2,3$ only, and in the plane defect geometry for $N=2,3,4$.

\subsection{Open BCs}
\label{sec:results:openbc}
\begin{figure*}
    \centering
    \includegraphics[width=0.9\linewidth]{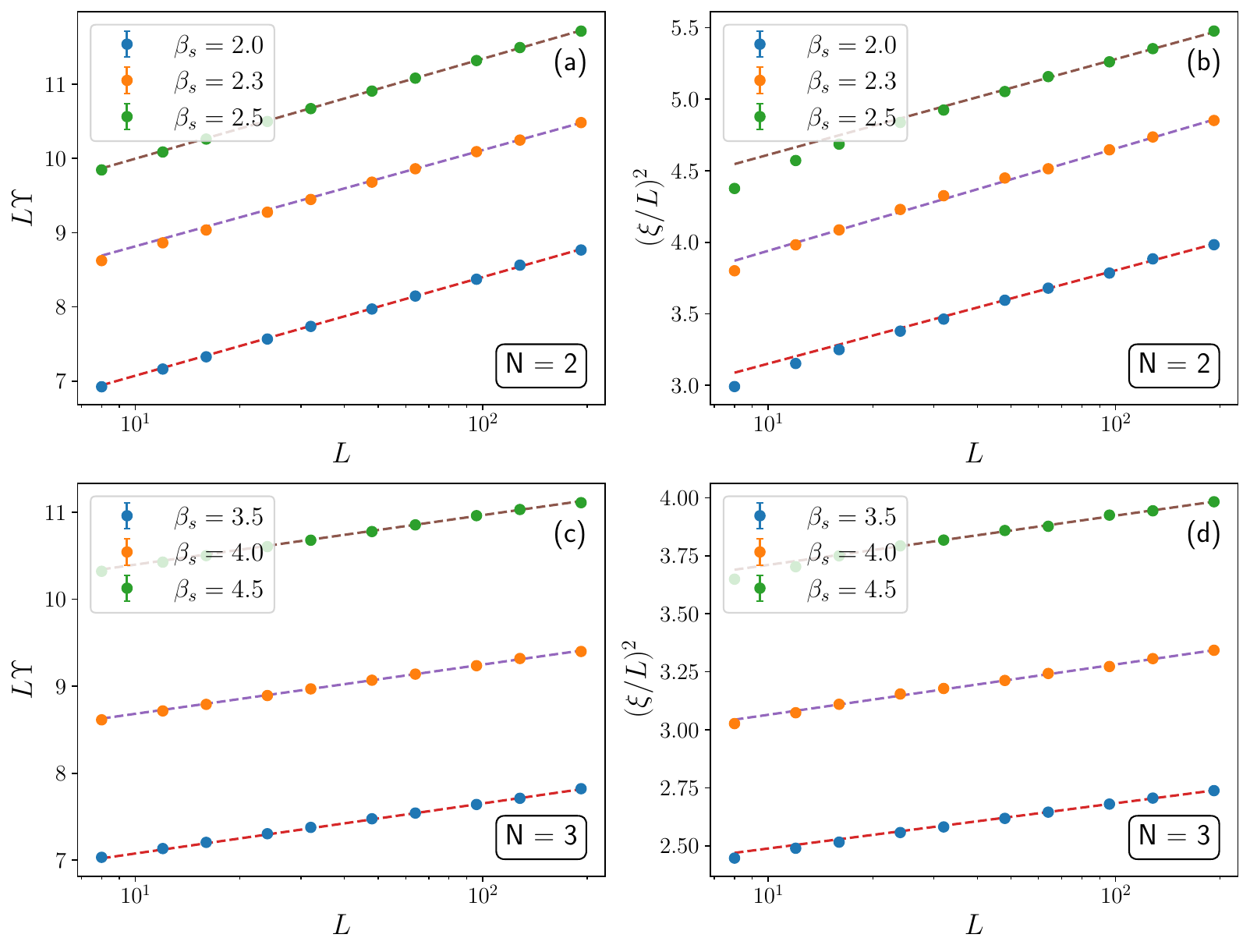}
    \caption{Rescaled helicity modulus $L\Upsilon$ and surface correlation length ratio $\xi/L$ for the $N=2,3$ model with open BCs and in the extraordinary-log phase.
    Dashed lines correspond to the fit to Eq.~(\ref{LY_fit_openbc}) (for $L\Upsilon$) and to Eq.~(\ref{xil_fit}) (for $\xi/L$).}
    \label{fig:LY_xil_openbc}
\end{figure*}

We have sampled the $N=2$, $3$ models with open BCs for a range of coupling constants $\beta_s$ in the extraordinary-log phase, and for lattice sizes $8\le L\le 192$.
In Figs \ref{fig:LY_xil_openbc}(a) and \ref{fig:LY_xil_openbc}(b) we plot the RG-invariants $L\Upsilon$ and $(\xi/L)^2$ for the XY model $N=2$ as a function of $L$ in a semilogarithmic scale.
In line with  Eq.~(\ref{FSSLY_open}) and Eq.~(\ref{FSSxil}), we observe a linear growth
as a function of $\ln L$.

\begin{table}[b]
    \centering
    \caption{Boundary geometry. Results of fits of $L\Upsilon$ to Eq.~(\ref{LY_fit_openbc}) (third and fourth column) and of $(\xi/L)^2$ to Eq.~(\ref{xil_fit}) (fifth and sixth column), for $N=2,3$ and open BCs for some values of the boundary coupling constants $\beta_s$ in the extraordinary-log phase.}
    \begin{ruledtabular}
    \begin{tabular}{cbabab}
         & &\multicolumn{2}{c}{$L\Upsilon$} & \multicolumn{2}{c}{$(\xi/L)^2$} \\
         \cline{3-4} \cline{5-6}
         \multicolumn{1}{c}{\rule{0pt}{2.5ex}$N$} & \multicolumn{1}{c}{$\beta_s$} & \multicolumn{1}{c}{$\alpha$} & \multicolumn{1}{c}{$\chi^2/\text{d.o.f.}$} & \multicolumn{1}{c}{$\alpha$} & \multicolumn{1}{c}{$\chi^2/\text{d.o.f.}$} \\
         \hline
        &  2.0  &  0.2884(69)  & 1.9  & 0.2833(78)  & 3.4 \\ 
      2 &  2.3  &  0.2814(71)  & 0.1  & 0.3106(97)  & 0.6 \\ 
        &  2.5  &  0.2923(72)  & 0.2  & 0.291(11)  & 1.6 \\ 
         \rule{0pt}{2.5ex}
        &  3.5  &  0.1875(62)  & 0.3  & 0.1773(58)  & 0.1 \\ 
      3 &  4.0  &  0.1844(61)  & 1.3  & 0.1891(68)  & 2.6 \\ 
        &  4.5  &  0.1853(70)  & 1.2  & 0.1805(85)  & 3.2 \\ 
    \end{tabular}
    \end{ruledtabular}
    \label{tab:fits_openbc}
\end{table}
For a more quantitative analysis, we fit $L\Upsilon$ and $(\xi/L)^2$ to
\begin{align}
    \label{LY_fit_openbc}
    L\Upsilon = L\Upsilon_0 + \left(4\alpha / N\right)\ln L,\\
    \label{xil_fit}
    (\xi/L)^2 = (\xi/L)^2_0 + \left(\alpha / (N-1)\right)\ln L,
\end{align}
with $N=2$ and
leaving $L\Upsilon_0$, $(\xi/L)^2_0$, and $\alpha$ as free parameters.
To reduce the influence of subleading corrections not included in Eqs.~(\ref{LY_fit_openbc}) and (\ref{xil_fit}), we include in the fits the MC data for the four largest lattices $L=64,96,128,192$;
a comparison of MC data with the fitted curves is shown
in Figs.~\ref{fig:LY_xil_openbc}(a) and \ref{fig:LY_xil_openbc}(b).
In Table \ref{tab:fits_openbc} we report the fitted values of $\alpha$.
We stress that the error bars given in Table \ref{tab:fits_openbc} are those obtained from the fitting procedure and likely underestimate the actual uncertainty in $\alpha$.
In fact, the next term $O(g^3)$ in the $\beta-$function (\ref{beta}) gives rise to corrections $\propto \ln \ln L$, which are unfeasible to a direct numerical analysis and hamper a robust determination of $\alpha$.
Moreover, the presence of subleading corrections is indirectly confirmed by
the large $\chi^2/\text{d.o.f.}$ (d.o.f. denotes the degrees of freedom of the fit) found in some fits.
Despite these caveats,
we observe a nice agreement with the value of $\alpha$  extracted from the normal UC, reported in Table \ref{tab:alpha}. A more detailed analysis that includes a quantitative determination of the next term $b g^3$ in the $\beta$-function (\ref{beta}) is presented in Sec.~\ref{sec:results:beta}.

\begin{figure*}
    \centering
    \includegraphics[width=0.9\linewidth]{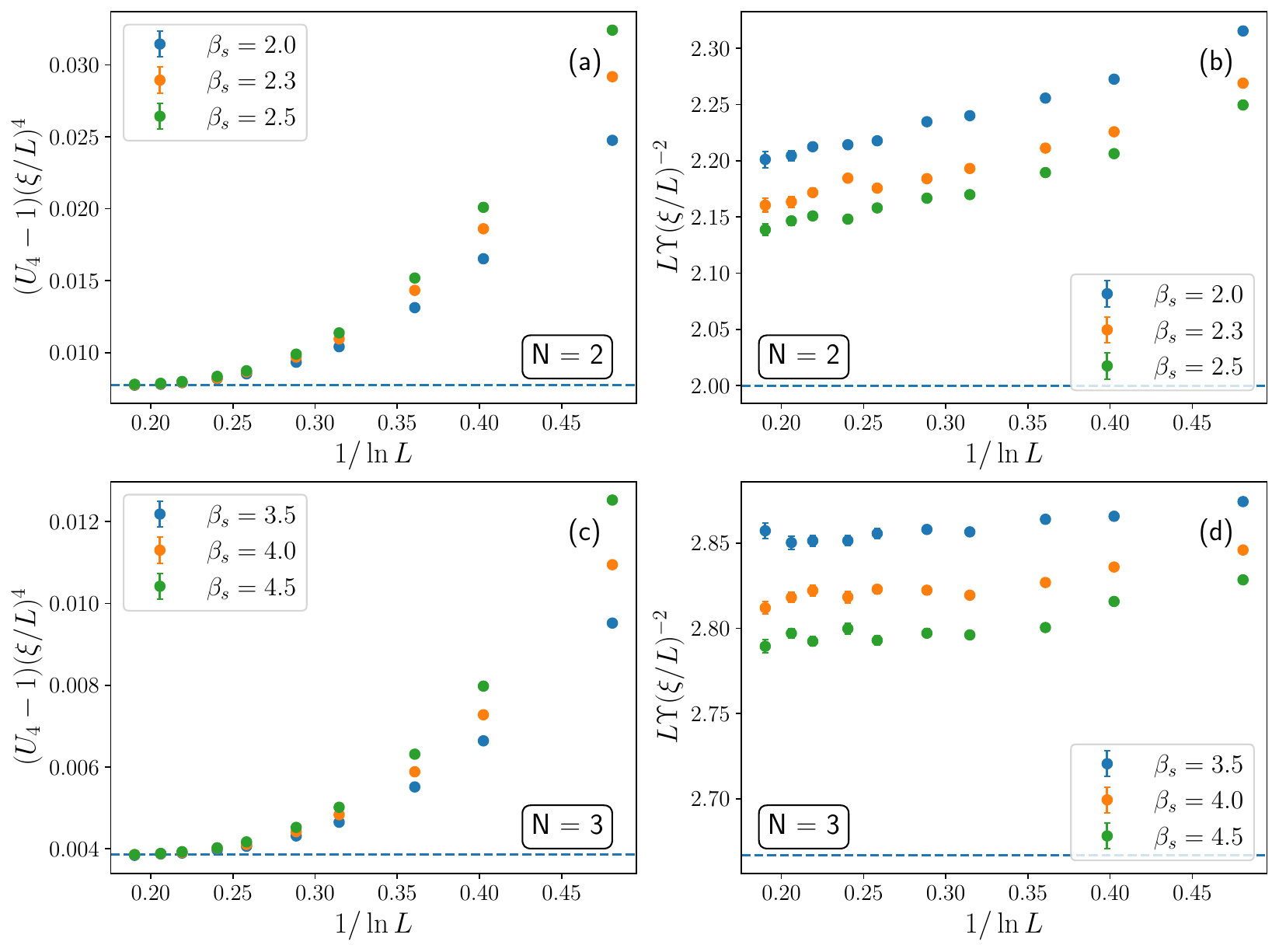}
    \caption{RG-invariant combinations $(U_4-1)(\xi/L)^4$ and $L\Upsilon(\xi/L)^{-2}$ for the $N=2,3$ model with open BCs and in the extraordinary-log phase.
   Dashed lines indicate the asymptotic value for $L\rightarrow \infty$ [see Eq.~(\ref{FSSxilU4}), and Eq.~(\ref{FSSLYxil_open})]: $(U_4-1)(\xi/L)^4 \to {\cal S}/(8 \pi^4) \simeq 0.00773$ and $L\Upsilon(\xi/L)^{-2} \to 2$ for  $N=2$;  $(U_4-1)(\xi/L)^4 \to {\cal S}/(16 \pi^4) \simeq 0.003877$ and $L\Upsilon(\xi/L)^{-2} \to 8/3$ for $N=3$.}
    \label{fig:RGcomb_openbc}
\end{figure*}

In Figs.~\ref{fig:RGcomb_openbc}(a) and \ref{fig:RGcomb_openbc}(b) we show the combinations of RG-invariants $(U_4-1)(\xi/L)^4$ and $L\Upsilon(\xi/L)^{-2}$ as a function of $1/\ln L$.
We observe a nice convergence of $(U_4-1)(\xi/L)^4$ to its limiting value given in  Eq.~(\ref{FSSxilU4}).
In the case of  $L\Upsilon(\xi/L)^{-2}$ we observe a somewhat slower approach to the limiting value 2, with a residual deviation of $\approx 7\%-10\%$ at $L=192$.
This slow approach to the limiting value can be well explained in terms of residual corrections to Eqs.~(\ref{FSSxilU4}) and (\ref{FSSLYxil_open}), which are of $O(g)$. As $g\rightarrow 0$ logarithmically in $L$, a convergence to the limiting value would require unfeasible large lattice sizes.

We carry out a similar analysis for the Heisenberg model $N=3$.
In Figs.~\ref{fig:LY_xil_openbc}(c) and \ref{fig:LY_xil_openbc}(d) we show the RG-invariants $L\Upsilon$ and $(\xi/L)^2$ in a semilogarithmic scale.
As before, they exhibit the expected linear growth in $\ln L$.
The fitted value of $\alpha$ reported in Table \ref{tab:fits_openbc} reveals a good agreement  with the value given in Table \ref{tab:alpha}.

In Figs.~\ref{fig:RGcomb_openbc}(c) and \ref{fig:RGcomb_openbc}(d) we show the combinations of RG-invariants $(\xi/L)^4(U_4-1)$ and $L\Upsilon(\xi/L)^{-2}$.
Similar to the $N=2$ case, we observe a clear convergence of $(U_4-1)(\xi/L)^4$ to its limiting value given in  Eq.~(\ref{FSSxilU4}), whereas for reasons analogous to the $N=2$ case $L\Upsilon(\xi/L)^{-2}$ displays a residual deviation of $5\%-7\%$ from the limiting value $8/3$.

\subsection{Plane defect}
\label{sec:results:plane}
\begin{figure*}
    \centering
    \includegraphics[width=0.9\linewidth]{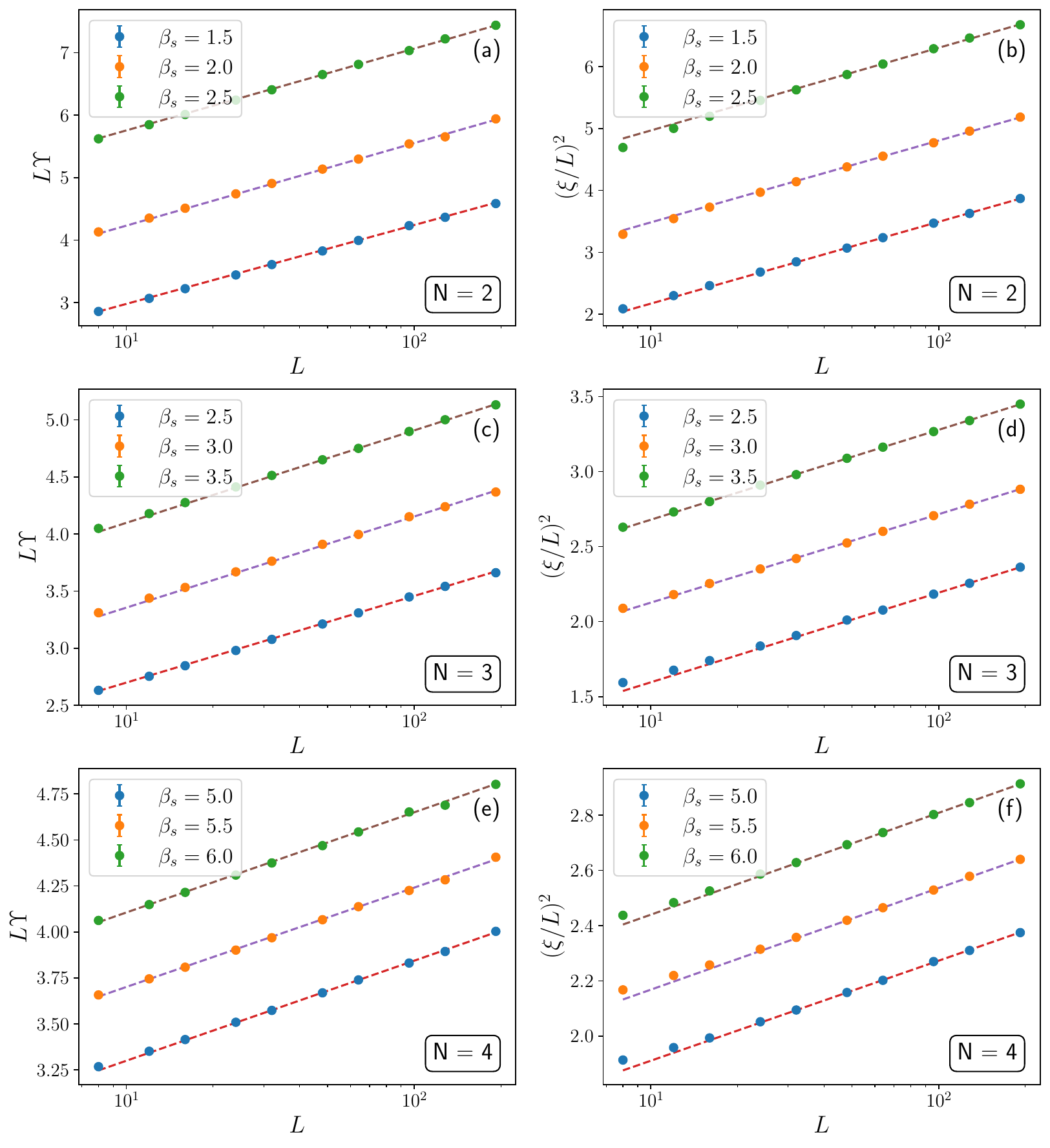}
    \caption{Rescaled helicity modulus $L\Upsilon$ and surface correlation length ratio $\xi/L$ for the $N=2,3,4$ model with  a plane defect in the extraordinary-log phase.
    Dashed lines correspond to the fit to Eq.~(\ref{LY_fit_plane}) (for $L\Upsilon$) and to Eq.~(\ref{xil_fit}) (for $\xi/L$).}
    \label{fig:LY_xil_plane}
\end{figure*}
We have simulated the model (\ref{model}) for $N=2,3,4$, lattice sizes $8\le L \le 192$, and in the presence of a plane defect.
As discussed in Sec.~\ref{sec:results:model}, we enhance the coupling constant on the plane, so as to realize the extraordinary-log UC. 
In Figs.~\ref{fig:LY_xil_plane} we plot the RG-invariant observables $L\Upsilon$ and $(\xi/L)^2$, which exhibit a linear growth in $\ln L$.

\begin{table}[b]
    \centering
    \caption{Plane defect geometry. Results of fits of $L\Upsilon$ to Eq.~(\ref{LY_fit_plane}) (third and fourth column) and of $(\xi/L)^2$ to Eq.~(\ref{xil_fit}) (fifth and sixth column), for $N=2,3,4$ for some values of the plane coupling constants $\beta_s$ in the extraordinary-log phase.}
    \begin{ruledtabular}
    \begin{tabular}{cbabab}
         & &\multicolumn{2}{c}{$L\Upsilon$} & \multicolumn{2}{c}{$(\xi/L)^2$} \\
         \cline{3-4} \cline{5-6}
         \multicolumn{1}{c}{\rule{0pt}{2.5ex}$N$} & \multicolumn{1}{c}{$\beta_s$} & \multicolumn{1}{c}{$\alpha$} & \multicolumn{1}{c}{$\chi^2/\text{d.o.f.}$} & \multicolumn{1}{c}{$\alpha$} & \multicolumn{1}{c}{$\chi^2/\text{d.o.f.}$} \\
         \hline
         & 1.5  &  0.548(18)  & 0.9  & 0.5738(74)  & 0.3 \\ 
       2 & 2.0  &  0.573(17)  & 2.0  & 0.5741(97)  & 2.1 \\ 
         & 2.5  &  0.568(16)  & 0.5  & 0.580(12)  & 1.0 \\ 
         \rule{0pt}{2.5ex}
         & 2.5  &  0.494(13)  & 1.0  & 0.5191(76)  & 0.1 \\ 
       3 & 3.0  &  0.520(15)  & 2.1  & 0.5122(72)  & 0.2 \\ 
         & 3.5  &  0.527(15)  & 0.4  & 0.5207(86)  & 0.2 \\ 
         \rule{0pt}{2.5ex}
         & 5.0  &  0.472(14)  & 0.7  & 0.4723(64)  & 2.1 \\ 
       4 & 5.5  &  0.469(15)  & 2.4  & 0.4798(74)  & 0.4 \\ 
         & 6.0  &  0.473(13)  & 3.7  & 0.4807(90)  & 0.2 \\ 
    \end{tabular}
    \end{ruledtabular}
    \label{tab:fits_plane}
\end{table}
In line with Eq.~(\ref{FSSLY_plane}), for this geometry we fit $L\Upsilon$ to
\begin{equation}
    L\Upsilon = L\Upsilon_0 + \left(2\alpha / N\right)\ln L,
    \label{LY_fit_plane}
\end{equation}
whereas for $(\xi/L)^2$ we use the Ansatz of Eq.~(\ref{xil_fit}), which is valid also for a plane defect.
Fit results are reported in Table \ref{tab:fits_plane}.
We generally observe, within error bars, a good agreement of the fitted values of $\alpha$ with the more precise determination of Table \ref{tab:alpha}.
Still,
especially for the smallest plane couplings $\beta_s$ and in the case $N=4$,
the fitted values of $\alpha$ appear to slightly underestimate the expected values of Table \ref{tab:alpha}, hinting at residual corrections. This aspect is further investigated in the following section.

\begin{figure*}
    \centering
    \includegraphics[width=0.9\linewidth]{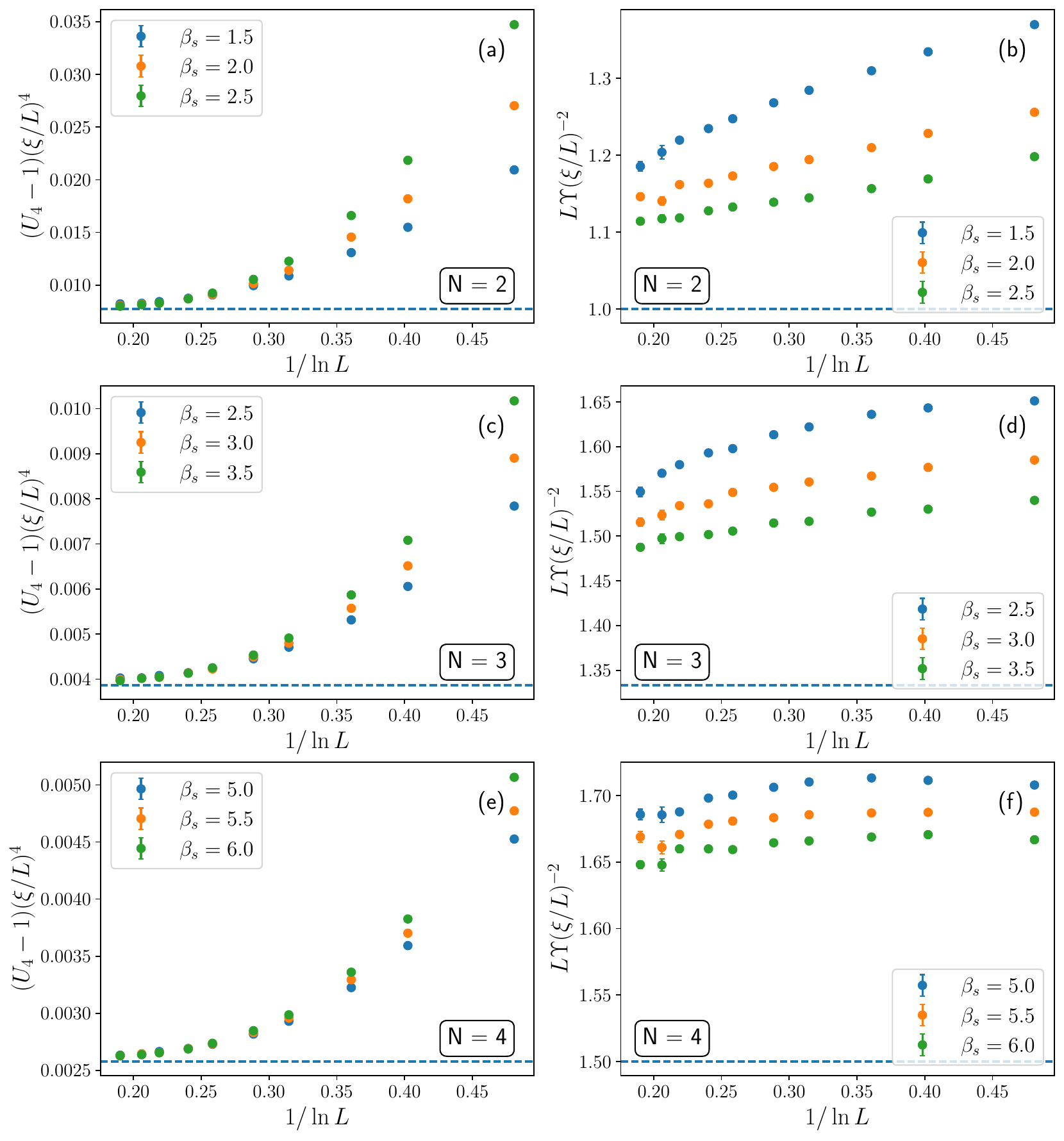}
    \caption{RG-invariant combinations $(\xi/L)^4(U_4-1)$ and $L\Upsilon(\xi/L)^{-2}$ for the $N=2,3,4$ model with plane defect in the extraordinary-log phase.
    Dashed lines indicate the asymptotic value for $L\rightarrow \infty$ [See Eq.~(\ref{FSSxilU4}), and Eq.~(\ref{FSSLYxil_plane})]: $(\xi/L)^4(U_4-1) \to {\cal S}/(8 \pi^4) \simeq 0.00773$ and $L\Upsilon(\xi/L)^{-2} \to 1$ for $N=2$;   $(\xi/L)^4(U_4-1) \to {\cal S}/(16 \pi^4) \simeq 0.003877$ and $L\Upsilon(\xi/L)^{-2} \to 4/3$ for  $N=3$;  $(\xi/L)^4(U_4-1) \to {\cal S}/(24 \pi^4) \simeq 0.002578$ and   $L\Upsilon(\xi/L)^{-2} \to 3/2$ for $N=4$.}
    \label{fig:RGcomb_plane}
\end{figure*}
In Fig.~\ref{fig:RGcomb_plane} we show the combinations of RG-invariants $(\xi/L)^4(U_4-1)$ and $L\Upsilon(\xi/L)^{-2}$.
Similar to the case of open BCs, $L\Upsilon(\xi/L)^{-2}$ converges to the expected $L\rightarrow\infty$  value slower than $(\xi/L)^4(U_4-1)$.

\subsection{\texorpdfstring{$\beta-$}{Beta-}function}
\label{sec:results:beta}
\begin{figure*}
    \centering
    \includegraphics[width=\linewidth]{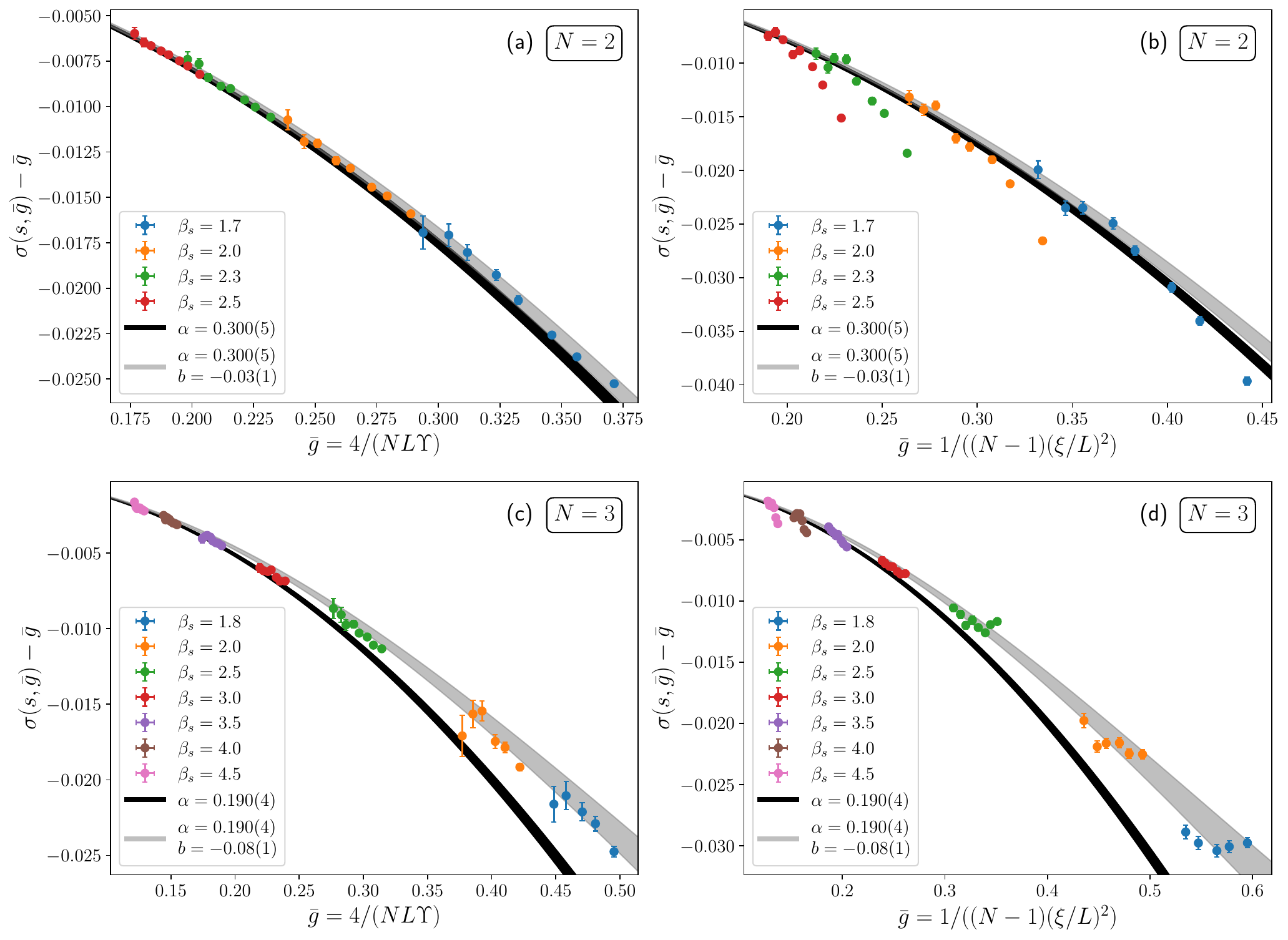}
    \caption{Step scaling function $\sigma(s, \bar{g})$ for open BCs and $s=2$. We plot $\sigma(s, \bar{g})$ computed for a range of surface coupling constant $\beta_s$ in the extraordinary-log phase,
    with the trivial linear term subtracted [See Eq.~(\ref{sigma})].
    (a),(c): $\sigma(s, \bar{g})$ for $N=2,3$ obtained with $\bar{g}$ defined in Eq.~(\ref{gLY_open}).
    (b),(d): $\sigma(s, \bar{g})$ for $N=2,3$ obtained with $\bar{g}$ defined in Eq.~(\ref{gxil}).
    For each value of $\beta_s$, points with bigger $\bar{g}$ correspond to the smallest lattices.
  As a comparison, in black and gray we plot $\sigma(s, \bar{g})$ truncated to the third order [see Eq.~(\ref{sigma})].
    The black range shows $\sigma(s, \bar{g})$ as obtained from the $\beta-$function truncated to the quadratic order, i.e., setting $b=0$ and employing the value of $\alpha$ reported in Table \ref{tab:alpha}.
    The gray range shows $\sigma(s, \bar{g})$ obtained with the value of $\alpha$ in Table \ref{tab:alpha} and our final estimate of $b$ given in Table \ref{tab:b}.
    The thickness of both curves reflects the uncertainty in the parameters $\alpha$ and $b$.
    }
    \label{fig:RGflow_open_s2}
\end{figure*}
In this subsection
we use our MC results to determine the coefficients of the $\beta-$function in Eq.~(\ref{beta}).
We follow closely the step-scaling-function method of Ref.~\cite{LWW-91}, which we summarize below.
Further, we benchmark this method using the 2D O$(3)$ as an example in Appendix \ref{app:beta_O32d}.

We define a renormalized coupling constant $\bar{g}$ in terms of the lattice data of $L\Upsilon$ and $(\xi/L)^2$.
In view of Eqs.~(\ref{FSSLY_open}), (\ref{FSSLY_plane}) and (\ref{FSSxil}), we choose the following definitions:
\begin{alignat}{2}
      \bar{g} \equiv \frac{4}{NL \Upsilon}, &\quad \text{open BCs},
      \label{gLY_open}\\
      \bar{g} \equiv \frac{2}{NL \Upsilon}, &\quad \text{plane defect},
      \label{gLY_plane}\\
      \bar{g} \equiv \frac{1}{(N-1)(\xi/L)^2}, &\quad \text{both geometries}.
      \label{gxil}
\end{alignat}
With these definitions $\bar{g} = g + O(g^2)$.
Given the choice of $\bar{g}$, and for fixed coupling constants, we introduce a step scaling function $\sigma(s, \bar{g})$ defined such that
\begin{equation}
    \sigma(s, \bar{g}(L)) \equiv \bar{g}(sL)
    \label{sigmadef}
\end{equation}
where $s$ is a parameter, here fixed to $s=2$; a separate analysis with $s=3/2$ is presented in Appendix \ref{app:beta_s1.5}.
The step scaling function describes the flow of the coupling constant $\bar{g}$ when the size is scaled as $L\rightarrow sL$.
  Since in a finite size system the RG scale is controlled by the system size $L$, $\sigma(s,\bar{g})$ is obtained from $\beta(\bar{g})$ with
\begin{equation}
    \ln s = - \int_{\bar{g}}^{\sigma(s,\bar{g})} \frac{du}{\beta(u)}.
  \end{equation}
Using the above equation, the Taylor expansion for $\bar{g}\rightarrow 0$ of $\sigma(s, \bar{g})$ can be directly related to the expansion of the $\beta(\bar{g})$ function as \cite{LWW-91}
\begin{align}
    \label{beta_gbar}
     \beta(\bar{g}) &= \alpha \bar{g}^2 + b\bar{g}^3 + O(\bar{g}^4),\\
    \label{sigma}
    \sigma(s,\bar{g}) &= \bar{g} - \left(\alpha \ln s\right)\bar{g}^2 + \left[\alpha^2\left(\ln s\right)^2 - b\ln s\right]\bar{g}^3 + O(\bar{g})^4.
\end{align}
Fitting  $\sigma(s,\bar{g})$ of the lattice model to
Eq.~(\ref{sigma}) provides a viable method to determine the coefficients of the $\beta-$function governing the RG-flow of $\bar{g}\sim g$.
We note that both $\alpha$ and $b$ are universal parameters that do not change on a redefinition of $g\rightarrow g + O(g)^2$.

\begin{figure*}
    \centering
    \includegraphics[width=\linewidth]{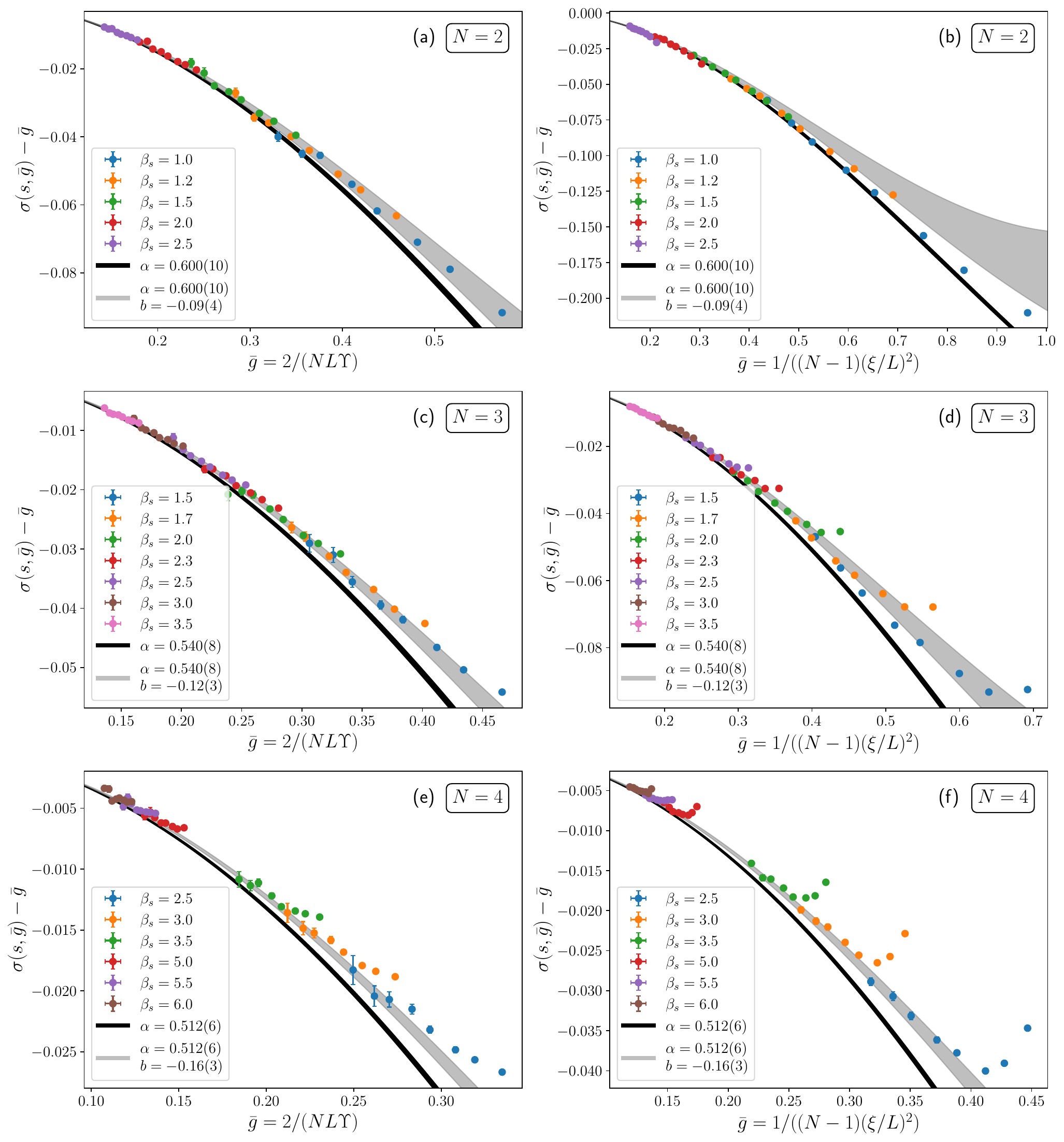}
    \caption{Same as Fig.~\ref{fig:RGflow_open_s2} for the plane-defect geometry and $N=2, 3, 4$.}
    \label{fig:RGflow_plane_s2}
\end{figure*}

\begin{table*}
    \centering
    \caption{Fits of $\sigma(s=2,\bar{g})$ to a third-order polynomial for open BCs and $N=2,3$.
    We report the resulting coefficients $\alpha$ and $b$ [See Eqs.~(\ref{beta_gbar}),(\ref{sigma})] as a function of maximum value $\bar{g}_{\rm max}$ of $\bar{g}$ and of the minimum lattice size $L_{\rm min}$ employed in the fit.
    In the left part of the Table,
    the renormalized coupling constant $\bar{g}$ is defined in terms of the helicity modulus [Eq.~(\ref{gLY_open}], in the right part we use the definition in terms of  $\xi / L$ [Eq.~(\ref{gxil})].
    }
    \begin{ruledtabular}
    \begin{tabular}{llaabllaab}
       \multicolumn{5}{c}{$\bar{g} = 4 / (N L\Upsilon)$}  & \multicolumn{5}{c}{$\bar{g} = 1 / [(N-1)(\xi/L)^2]$} \\
       \cline{1-5} \cline{6-10}
       \multicolumn{1}{l}{\rule{0pt}{2.5ex}$\bar{g}_{\rm max}$} & \multicolumn{1}{l}{$L_{\rm min}$} & \multicolumn{1}{c}{$\alpha$}  & \multicolumn{1}{c}{$b$} & \multicolumn{1}{c}{$\chi^2/\text{d.o.f.}$} &
       \multicolumn{1}{l}{$\bar{g}_{\rm max}$} & \multicolumn{1}{l}{$L_{\rm min}$} & \multicolumn{1}{c}{$\alpha$}  & \multicolumn{1}{c}{$b$} & \multicolumn{1}{c}{$\chi^2/\text{d.o.f.}$}\\
       \hline
       \multicolumn{10}{c}{\rule{0pt}{2.5ex}$N=2$} \\
       \hline\rule{0pt}{2.5ex}
       &   8  &   0.3092(16)  &  -0.0499(57)  & 1.5    &        &   8  &   0.3732(48)  &   -0.130(15)  & 46.1 \\
       &  12  &   0.3053(20)  &  -0.0392(72)  & 1.0    &        &  12  &   0.3573(60)  &   -0.137(20)  & 9.7 \\
  0.35 &  16  &   0.3034(25)  &  -0.0339(86)  & 0.9    &   0.35 &  16  &   0.3334(81)  &   -0.088(27)  & 5.0 \\
       &  24  &   0.3051(39)  &   -0.047(14)  & 0.5    &        &  24  &    0.316(10)  &   -0.054(34)  & 2.6 \\
       &  32  &   0.3089(65)  &   -0.061(24)  & 0.6    &        &  32  &    0.307(13)  &   -0.034(43)  & 2.9 \\[1em]

       &   8  &   0.3090(19)  &  -0.0494(71)  & 1.5    &        &   8  &   0.4165(87)  &   -0.292(31)  & 46.4 \\
       &  12  &   0.3040(28)  &   -0.034(11)  & 0.9    &        &  12  &   0.3855(93)  &   -0.249(34)  & 10.6 \\
   0.3 &  16  &   0.2983(44)  &   -0.012(18)  & 0.7    &    0.3 &  16  &    0.350(10)  &   -0.154(38)  & 5.2 \\
       &  24  &   0.3020(65)  &   -0.033(27)  & 0.5    &        &  24  &    0.318(12)  &   -0.062(41)  & 2.8 \\
       &  32  &    0.296(11)  &   -0.003(46)  & 0.5    &        &  32  &    0.313(17)  &   -0.057(59)  & 3.2 \\[1em]

       &   8  &   0.2975(48)  &    0.001(21)  & 1.3    &        &   8  &    0.374(25)  &    -0.14(10)  & 46.7 \\
       &  12  &   0.2929(71)  &    0.016(31)  & 0.6    &        &  12  &    0.393(24)  &    -0.29(10)  & 10.3 \\
  0.25 &  16  &    0.295(11)  &    0.003(49)  & 0.6    &   0.25 &  16  &    0.326(26)  &    -0.05(11)  & 5.8 \\
       &  24  &    0.306(16)  &   -0.053(72)  & 0.5    &        &  24  &    0.353(35)  &    -0.21(14)  & 2.2 \\
       &  32  &    0.296(22)  &  -0.005(102)  & 0.6    &        &  32  &    0.427(51)  &    -0.54(21)  & 1.8 \\
       \multicolumn{10}{c}{\rule{0pt}{2.5ex}$N=3$} \\[0.5em]
       \hline\rule{0pt}{2.5ex}
       &   8  &  0.20631(80)  &  -0.1022(35)  & 1.6    &        &   8  &   0.2263(15)  &  -0.1821(48)  & 10.6 \\
       &  12  &   0.2035(12)  &  -0.0910(52)  & 1.3    &        &  12  &   0.2112(18)  &  -0.1319(60)  & 4.1 \\
   0.4 &  16  &   0.2039(16)  &  -0.0979(73)  & 1.0    &    0.4 &  16  &   0.1986(24)  &  -0.0904(79)  & 1.3 \\
       &  24  &   0.2007(24)  &   -0.090(10)  & 0.8    &        &  24  &   0.1915(30)  &   -0.068(10)  & 0.9 \\
       &  32  &   0.2014(33)  &   -0.096(14)  & 0.8    &        &  32  &   0.1948(42)  &   -0.079(14)  & 0.9 \\[1em]

       &   8  &   0.2074(12)  &  -0.1086(59)  & 1.5    &        &   8  &   0.2361(26)  &   -0.234(12)  & 10.7 \\
       &  12  &   0.2035(16)  &  -0.0914(79)  & 1.1    &        &  12  &   0.2157(34)  &   -0.156(16)  & 4.4 \\
   0.3 &  16  &   0.2027(20)  &  -0.0915(99)  & 1.0    &    0.3 &  16  &   0.1971(43)  &   -0.083(21)  & 1.3 \\
       &  24  &   0.1994(27)  &   -0.083(12)  & 0.7    &        &  24  &   0.1920(58)  &   -0.070(29)  & 0.8 \\
       &  32  &   0.1999(40)  &   -0.088(19)  & 0.7    &        &  32  &   0.1929(78)  &   -0.069(38)  & 0.8 \\[1em]

       &   8  &   0.2095(21)  &   -0.122(13)  & 1.6    &        &   8  &   0.2638(59)  &   -0.404(34)  & 13.6 \\
       &  12  &   0.2068(31)  &   -0.113(19)  & 1.1    &        &  12  &   0.2313(65)  &   -0.251(38)  & 5.7 \\
   0.2 &  16  &   0.2050(44)  &   -0.106(27)  & 1.1    &    0.2 &  16  &   0.1987(76)  &   -0.092(43)  & 1.7 \\
       &  24  &   0.2026(69)  &   -0.103(43)  & 0.8    &        &  24  &   0.2007(97)  &   -0.122(55)  & 0.9 \\
       &  32  &    0.211(10)  &   -0.164(65)  & 0.7    &        &  32  &    0.198(14)  &   -0.102(77)  & 1.0 \\
    \end{tabular}
    \end{ruledtabular}
    \label{tab:fits_RGflow_open_s2}
\end{table*}
In Fig.~\ref{fig:RGflow_open_s2} and Fig.~\ref{fig:RGflow_plane_s2} we show the step scaling function $\sigma(s=2, \bar{g})$, with the trivial linear term $\bar{g}$ subtracted [see Eq.~(\ref{sigma})], obtained for both choices of $\bar{g}$ given in Eqs.~(\ref{gLY_open})-(\ref{gxil}) and for all models considered here.
The scaling function is computed using the MC data presented in Secs.~\ref{sec:results:openbc} and \ref{sec:results:plane}, as well as employing results of simulations for some additional values of the boundary coupling in the extraordinary-log phase, and for lattice sizes $8\le L\le 192$.
Inspecting Fig.~\ref{fig:RGflow_open_s2} and Fig.~\ref{fig:RGflow_plane_s2} we observe that on an optical scale the data obtained from the helicity modulus appear to align to a single curve.
In the case of $\bar{g}$ defined in terms of $\xi/L$, such a collapse is found only for larger lattice sizes, with significant deviations observed for smaller lattice sizes,
thus suggesting important finite-size corrections.
In Fig.~\ref{fig:RGflow_open_s2} and Fig.~\ref{fig:RGflow_plane_s2} we also plot (black range)  $\sigma(s=2, \bar{g})-\bar{g}$ truncated to the third order [as in the right-hand side of Eq.~(\ref{sigma})],
 setting $\alpha$ to the expected value in Table \ref{tab:alpha} and  $b =0$.
The resulting curve matches well the MC data at small $\bar{g}$, whereas clear deviations are visible for larger values of $\bar{g}$.
Therefore, in the range of MC data shown in Fig.~\ref{fig:RGflow_open_s2} and Fig.~\ref{fig:RGflow_plane_s2}, the $b g^3$  correction to the $\beta-$function is numerically
 important to describe the RG flow.

For a quantitative determination of the $\beta-$function, we have fitted the MC data of $\sigma(\bar{g})$ to the right-hand side of Eq.~(\ref{sigma}), fixing $s=2$ and leaving $\alpha$ and $b$ as free parameters \footnote{We notice that the MC data for these fits are statistically correlated because, for each $L$, the pair of data points $[\bar{g}(L), \sigma(s, \bar{g}(L))]$ and $[\bar{g}(sL), \sigma(s, \bar{g}(sL))]$ is constructed using the same MC result for $\bar{g}(L)$.
Furthermore, data points are affected by a statistical error bar on both coordinates.
Our fits take these technical aspects into account, by employing the full covariance matrix in the $\chi^2$ cost function.}.
Fit results are shown in Table \ref{tab:fits_RGflow_open_s2}, for the boundary geometry, and in Table \ref{tab:fits_RGflow_plane_s2} for the plane-defect geometry.
In order to monitor the influence of corrections to the polynomial expansion, we have done the fits using MC data for $\bar{g} < \bar{g}_{\rm max}$, for a few values of the cutoff $\bar{g}_{\rm max}$.
We notice that the analysis of Eqs.~(\ref{beta_gbar}) and (\ref{sigma}) neglects corrections from irrelevant operators, and from nonuniversal effects,
which are especially evident in the MC data for $\xi/L$ [see Figs.~\ref{fig:RGflow_open_s2}(b), ~\ref{fig:RGflow_open_s2}(d) and \ref{fig:RGflow_plane_s2}(b), \ref{fig:RGflow_plane_s2}(d), \ref{fig:RGflow_plane_s2}(f)].
A detailed study of such corrections is beyond the scope of this work.
Nevertheless, in order to attempt an extrapolation to the large-$L$ limit, and to monitor
the relevance of finite-size corrections in the final results, for every value of the cutoff $\bar{g}_{\rm max}$, we have systematically discarded the smallest lattice sizes in the fits.

\begin{table*}
    \centering
    \caption{Same as Table \ref{tab:fits_RGflow_open_s2} for the plane-defect geometry.}
    \begin{ruledtabular}
    \begin{tabular}{llaabllaab}
       \multicolumn{5}{c}{$\bar{g} = 4 / (N L\Upsilon)$}  & \multicolumn{5}{c}{$\bar{g} = 1 / [(N-1)(\xi/L)^2]$} \\
       \cline{1-5} \cline{6-10}
       \multicolumn{1}{l}{\rule{0pt}{2.5ex}$\bar{g}_{\rm max}$} & \multicolumn{1}{l}{$L_{\rm min}$} & \multicolumn{1}{c}{$\alpha$}  & \multicolumn{1}{c}{$b$} & \multicolumn{1}{c}{$\chi^2/\text{d.o.f.}$} &
       \multicolumn{1}{l}{$\bar{g}_{\rm max}$} & \multicolumn{1}{l}{$L_{\rm min}$} & \multicolumn{1}{c}{$\alpha$}  & \multicolumn{1}{c}{$b$} & \multicolumn{1}{c}{$\chi^2/\text{d.o.f.}$}\\
       \hline
       \multicolumn{10}{c}{\rule{0pt}{2.5ex}$N=2$} \\
       \hline\rule{0pt}{2.5ex} 
       &   8  &   0.5843(18)  &  -0.0749(53)  & 5.8    &        &   8  &   0.6282(21)  &  -0.0657(33)  & 23.0 \\
       &  12  &   0.5861(26)  &  -0.0582(76)  & 2.1    &        &  12  &   0.6049(24)  &  -0.0325(37)  & 7.5 \\
   0.4 &  16  &   0.5880(35)  &  -0.0531(94)  & 1.4    &    0.6 &  16  &   0.5943(28)  &  -0.0168(40)  & 3.9 \\
       &  24  &   0.5914(58)  &   -0.057(16)  & 1.6    &        &  24  &   0.5856(33)  &   0.0002(50)  & 2.0 \\
       &  32  &   0.5800(91)  &   -0.024(24)  & 1.7    &        &  32  &   0.5824(44)  &   0.0067(60)  & 1.7 \\[1em]

       &   8  &   0.5937(39)  &   -0.115(16)  & 3.4    &        &   8  &   0.6630(32)  &  -0.1481(69)  & 16.4 \\
       &  12  &   0.5919(56)  &   -0.082(23)  & 1.1    &        &  12  &   0.6335(36)  &  -0.1013(74)  & 3.1 \\
   0.3 &  16  &   0.5933(72)  &   -0.078(29)  & 1.0    &   0.45 &  16  &   0.6196(47)  &   -0.080(10)  & 1.3 \\
       &  24  &   0.6022(91)  &   -0.107(34)  & 1.0    &        &  24  &   0.6096(59)  &   -0.062(14)  & 0.8 \\
       &  32  &    0.595(15)  &   -0.096(59)  & 1.0    &        &  32  &   0.6061(80)  &   -0.055(19)  & 0.9 \\[1em]

       &   8  &    0.692(27)  &    -0.60(13)  & 1.5    &        &   8  &   0.6739(86)  &   -0.200(29)  & 22.9 \\
       &  12  &    0.631(42)  &    -0.28(22)  & 1.3    &        &  12  &   0.6210(92)  &   -0.054(31)  & 4.8 \\
   0.2 &  16  &    0.615(48)  &    -0.19(25)  & 1.4    &    0.3 &  16  &    0.609(12)  &   -0.041(42)  & 2.0 \\
       &  24  &    0.591(50)  &    -0.04(27)  & 1.2    &        &  24  &    0.608(15)  &   -0.057(56)  & 1.1 \\
       &  32  &    0.580(52)  &  -0.003(277)  & 1.3    &        &  32  &    0.599(20)  &   -0.029(76)  & 1.2 \\
       \multicolumn{10}{c}{\rule{0pt}{2.5ex}$N=3$} \\
       \hline\rule{0pt}{2.5ex}
       &   8  &   0.5263(15)  &  -0.1320(49)  & 10.0   &        &   8  &   0.5280(12)  &  -0.1188(20)  & 69.1 \\
       &  12  &   0.5358(21)  &  -0.1321(65)  & 1.8    &        &  12  &   0.5288(14)  &  -0.0924(24)  & 14.8 \\
   0.4 &  16  &   0.5379(31)  &  -0.1286(94)  & 1.4    &    0.6 &  16  &   0.5320(18)  &  -0.0865(28)  & 7.5 \\
       &  24  &   0.5401(45)  &   -0.122(13)  & 0.8    &        &  24  &   0.5315(24)  &  -0.0711(42)  & 2.1 \\
       &  32  &   0.5334(72)  &   -0.103(23)  & 0.8    &        &  32  &   0.5309(36)  &  -0.0600(69)  & 1.4 \\[1em]

       &   8  &   0.5345(23)  &  -0.1665(89)  & 9.0    &        &   8  &   0.5530(19)  &  -0.1899(44)  & 30.8 \\
       &  12  &   0.5397(37)  &   -0.148(14)  & 1.8    &        &  12  &   0.5390(24)  &  -0.1218(59)  & 6.7 \\
   0.3 &  16  &   0.5385(51)  &   -0.132(20)  & 1.4    &   0.45 &  16  &   0.5351(31)  &  -0.0973(77)  & 4.2 \\
       &  24  &   0.5404(73)  &   -0.124(28)  & 0.9    &        &  24  &   0.5340(37)  &  -0.0786(92)  & 1.4 \\
       &  32  &    0.533(11)  &   -0.100(44)  & 1.0    &        &  32  &   0.5382(47)  &   -0.081(11)  & 1.2 \\[1em]

       &   8  &   0.5120(99)  &   -0.041(52)  & 6.1    &        &   8  &   0.5447(44)  &   -0.152(16)  & 4.4 \\
       &  12  &    0.546(11)  &   -0.185(55)  & 1.6    &        &  12  &   0.5503(49)  &   -0.161(17)  & 2.9 \\
   0.2 &  16  &    0.568(16)  &   -0.290(82)  & 1.4    &    0.3 &  16  &   0.5418(64)  &   -0.118(23)  & 2.2 \\
       &  24  &    0.530(27)  &    -0.06(14)  & 1.1    &        &  24  &   0.5345(77)  &   -0.078(28)  & 1.2 \\
       &  32  &    0.584(42)  &    -0.39(23)  & 0.8    &        &  32  &   0.5377(98)  &   -0.080(34)  & 1.2 \\
       \multicolumn{10}{c}{\rule{0pt}{2.5ex}$N=4$} \\
       \hline\rule{0pt}{2.5ex}
       &   8  &   0.4858(14)  &  -0.2268(65)  & 16.5   &        &   8  &   0.4878(17)  &  -0.2549(56)  & 64.4 \\
       &  12  &   0.4942(21)  &  -0.2079(96)  & 3.3    &        &  12  &   0.4902(22)  &  -0.1990(73)  & 15.2 \\
   0.3 &  16  &   0.4961(30)  &   -0.190(14)  & 1.8    &   0.35 &  16  &   0.4913(28)  &  -0.1642(96)  & 4.8 \\
       &  24  &   0.5036(43)  &   -0.202(18)  & 1.4    &        &  24  &   0.4926(34)  &   -0.146(12)  & 1.3 \\
       &  32  &   0.5142(70)  &   -0.235(30)  & 1.2    &        &  32  &   0.4943(48)  &   -0.139(17)  & 1.1 \\[1em]

       &   8  &   0.4845(19)  &   -0.219(10)  & 15.7   &        &   8  &   0.4547(49)  &   -0.082(28)  & 25.8 \\
       &  12  &   0.4912(28)  &   -0.188(15)  & 3.0    &        &  12  &   0.4697(52)  &   -0.082(29)  & 5.9 \\
  0.25 &  16  &   0.4921(39)  &   -0.164(20)  & 1.8    &   0.25 &  16  &   0.4801(56)  &   -0.100(30)  & 2.3 \\
       &  24  &   0.5002(53)  &   -0.182(26)  & 1.4    &        &  24  &   0.4902(60)  &   -0.130(31)  & 0.8 \\
       &  32  &   0.5133(87)  &   -0.231(43)  & 1.2    &        &  32  &   0.4982(72)  &   -0.160(34)  & 0.6 \\[1em]

       &   8  &   0.5093(61)  &   -0.390(41)  & 13.5   &        &   8  &    0.511(30)  &    -0.40(20)  & 13.4 \\
       &  12  &   0.4947(84)  &   -0.213(57)  & 2.6    &        &  12  &    0.536(30)  &    -0.49(20)  & 2.2 \\
   0.2 &  16  &    0.495(11)  &   -0.185(76)  & 2.1    &   0.15 &  16  &    0.550(31)  &    -0.58(20)  & 2.0 \\
       &  24  &    0.500(13)  &   -0.187(93)  & 1.8    &        &  24  &    0.520(37)  &    -0.34(25)  & 0.8 \\
       &  32  &    0.516(16)  &    -0.25(10)  & 1.5    &        &  32  &    0.526(49)  &    -0.35(34)  & 0.6 \\    
    \end{tabular}
    \end{ruledtabular}
    \label{tab:fits_RGflow_plane_s2}
\end{table*}

\begin{table*}[t]
    \centering
    \caption{Comparison of the value of the subleading coefficient $b$ in the $\beta$-function (\ref{beta}) obtained in Sec.~\ref{sec:results:beta} from FSS analysis (superscript MC) and from leading  large $N$ approximation (superscript $N$) for open and defect geometries. For the open BC, we also list the value of $b$ ( $b^{\Delta}_{{\rm OBC}}$,  $b^{\nu}_{{\rm OBC}}$,  $b^{U}_{{\rm OBC}}$) imputed from the critical properties at the special transition in Table \ref{tab:sp}, by assuming the perturbative results in Eqs.~(\ref{Deltan}) and (\ref{Ucr}).
    }
    \begin{ruledtabular}
    \begin{tabular}{laaaafae}
    \multicolumn{1}{l}{$N$} & \multicolumn{1}{c}{$b^{\rm MC}_{{\rm OBC}}$} & \multicolumn{1}{c}{$b^{N}_{{\rm OBC}}$} &
    \multicolumn{1}{c}{$b^{\Delta}_{{\rm OBC}}$} & \multicolumn{1}{c}{$b^{\nu}_{{\rm OBC}}$} & \multicolumn{1}{c}{$b^{U}_{{\rm OBC}}$} &
    \multicolumn{1}{c}{$b^{\rm MC}_{{\rm def}}$} & \multicolumn{1}{c}{$b^{N}_{{\rm def}}$} \\[0.5em]
    \hline
       2     & -0.03 (1) & -0.068 & -0.073(1) & -0.148(5)  & -0.060(1) & -0.09(4)  & -0.084  \\
       3     & -0.08 (1) & -0.101 & -0.115(3) & -0.100(5)  & -0.093(2) & -0.12(3)  & -0.127  \\
       4     &           & -0.135 & -0.126(4) & -0.088(14) & -0.111(4) &  -0.16(3) & -0.169  \\
    \end{tabular}
    \end{ruledtabular}
    \label{tab:b}
\end{table*}

In Table \ref{tab:fits_RGflow_open_s2} we show fit results for open BCs, and in Table \ref{tab:fits_RGflow_plane_s2} for the plane-defect geometry.
At $N=2$, fits for $L\Upsilon$ deliver a value of $\alpha$ perfectly consistent with the known estimate. The correction term on the other hand appears to be small, and we can roughly estimate $b\approx -0.03(1)$.
Fits for $\xi/L$ are of worse quality, displaying a significantly large $\chi^2/\text{d.o.f.}$, consistent with evident corrections to the leading behavior seen in Fig~\ref{fig:RGflow_open_s2}. Nevertheless, the fitted value of $\alpha$
obtained using only the largest lattice sizes is overall consistent within one error bar with the expected value of Table \ref{tab:alpha}.
For $N=3$, the fits of $L\Upsilon$ give a value of $\alpha$ mostly in marginal, in a few cases full, agreement with $\alpha=0.190(4)$ \cite{PTM-21}.
The fitted coefficient $b$ is small and negative; considering the fits where $\alpha$ is closer to the expected value,
we can estimate $b\simeq -0.08(1)$.
For $N=3$ fits for $\xi/L$ give a $\alpha$ perfectly consistent with previous estimates, and we can estimate $b\simeq -0.08(2)$, fully consistent with the estimate obtained with $L\Upsilon$.

In the case of plane-defect geometry, we generically observe larger coefficients $b$.
For $N=2$, fits exhibit a $\chi^2/\text{d.o.f.}$ which is larger than for the open BCs case.
Despite this difficulty, even for fits with a significant large $\chi^2/\text{d.o.f.}$ the fitted value of $\alpha$ obtained with $\bar{g}$ defined in terms of the helicity modulus agrees within uncertainty with the expected value $\alpha=0.600(10)$.
A good $\chi^2/\text{d.o.f.}$ is obtained for $\bar{g}_{\rm max}=0.3$ and $L_{\rm min}\ge 16$.
Based on the resulting value of $b$ for these fits, we can roughly estimate $b=-0.09(4)$.
A qualitatively similar picture is found for $\bar{g}$ defined in Eq.~(\ref{gxil}): despite a moderately large $\chi^2/\text{d.o.f.}$, for $L_{\rm min}\ge 16$ the fitted $\alpha$ agrees within uncertainty with the value of Table \ref{tab:alpha}.
Here we can estimate $b=-0.06(2)$, which agrees within error bars with the estimate extracted from the analysis of $L\Upsilon$.
For $N=3$ and $\bar{g}$ defined from the helicity modulus, fits are rather stable, and deliver a value of $\alpha$ in agreement with the value reported in Table \ref{tab:alpha}, $\alpha=0.540(8)$.
The correction term $b$ can be estimated as $b=-0.12(3)$.
In the case of $\bar{g}$ defined in terms of $\xi/L$, fits exhibit a worse $\chi^2/\text{d.o.f.}$.
Again, except for smaller lattices, the fitted value of $\alpha$ agrees within error bars with that of Table \ref{tab:alpha}.
From the fits with $L_{\rm min}=32$ we can roughly estimate $b=-0.08(3)$, though also in this case we find a somewhat large residual $\chi^2/\text{d.o.f.}\approx 1.2$.
Nonetheless, such fitted value of $b$ agrees within error bars with the more stable result obtained using $L\Upsilon$.
For $N=4$ and $\bar{g}$ defined from the helicity modulus, fits show a moderately large $\chi^2/\text{d.o.f.}$, which decreases to $\chi^2/\text{d.o.f.}\approx 1.2$ for $L_{\rm min}=32$.
Nevertheless,
within one error bar
the fitted value of $\alpha$ for $L_{\rm min}=24$ agrees with the expected value of Table \ref{tab:alpha}; a full agreement is found for $L_{\rm min}=32$.
Considering the fits with the largest lattices $L\ge L_{\rm min}=32$, one can estimate $b\simeq -0.23(4)$.
For $\bar{g}$ defined in terms of $\xi/L$, fits show a good $\chi^2/\text{d.o.f.}$ for the largest lattices.
The fitted value of $\alpha$ exhibits however a small deviation from the value of Table \ref{tab:alpha}.
The coefficient $b$ can be estimated as $b\simeq -0.16(3)$. This value is also in marginal agreement with the estimate obtained from the less satisfactory fits of $L\Upsilon$.

In line with the considerations outlined above, we give in Table \ref{tab:b} our estimates of the coefficient $b$ of the $\beta-$function in the column $b^{\rm MC}_{\rm OBC}$ (open BCs) and $b^{\rm MC}_{\rm def}$ (defect).
In Figs.~\ref{fig:RGflow_open_s2} and \ref{fig:RGflow_plane_s2} we also plot (gray range) the resulting step function  obtained employing such estimates, together with the values of $\alpha$ given in Table \ref{tab:alpha}.
Compared to the truncation of the $\beta-$function to the $O(g^2)$ order (black range), the resulting curve matches
rather well the MC data in a wider interval of $g$, underscoring the relevance of the $bg^3$ term in the $\beta-$function for a numerically accurate characterization of the RG flow.

In all cases studied here, we have observed a negative coefficient $b$.
Therefore, the $g^3$ term in the $\beta-$function has the effect of slowing the RG flow of $g\rightarrow 0$.
This implies that, in a scaling analysis which considers only the {\it leading} term $\alpha g^2$ of the $\beta-$function, the effectively observed RG-parameter $\alpha$ underestimates its actual value.
This may explain why some of the fits to the logarithmic behavior of $L\Upsilon$ and $\xi/L$ reported in Sec.~\ref{sec:results:openbc} and Sec.~\ref{sec:results:plane} give a value of $\alpha$ slightly below the expected values of Table \ref{tab:alpha}.
The correction term $b$, and hence such bias, increases in magnitude from $N=2$ to $N=4$;
this in particular  may provide an explanation of the residual small difference of the fitted $\alpha$ for $N=4$ plane defect geometry given in Table \ref{tab:fits_plane} and the value reported in Table \ref{tab:alpha}.

\section{Summary and Outlook: the evolution of the phase diagram with \texorpdfstring{$N$}{N} and the special transition}
\label{sec:outlook}
\begin{table}[b]
    \centering
    \caption{Critical properties at the special transition.
    }
    \begin{ruledtabular}
    \begin{tabular}{laaa}
    \multicolumn{1}{l}{$N$} & \multicolumn{1}{c}{$\Delta_{\vec{n}}$} & \multicolumn{1}{c}{$\nu^{-1}$} & \multicolumn{1}{c}{$U_4$} \\[0.5em]
    \hline
      2 \cite{DBN-05}     & 0.325 (1) & 0.608(4)  & 1.191(1)   \\
      3 \cite{PT-20}      & 0.264 (1) & 0.36(1)   &  1.0652(5) \\
      4 \cite{Deng-06}    & 0.184(2)  & 0.107(15) & 1.0178(8)  \\
    \end{tabular}
    \end{ruledtabular}
    \label{tab:sp}
\end{table}

In this work we have studied the FSS behavior of the extraordinary-log phase, which is found at the bidimensional surface and at a defect plane of a 3D O($N$) model.
This behavior is determined by the $\beta-$function governing the RG-flow at the boundary/defect, giving rise to a logarithmic dependence of RG-invariant observables on the size of the system.
Using MC simulations and a FSS analysis we have elucidated such a behavior, and computed the leading term $\alpha$ and the subleading one $b$ in the expansion of the $\beta-$function in Eq.~(\ref{beta}).
In line with theoretical predictions of Refs.~\cite{Metlitski-20,KM-23}, the value of $\alpha$ matches very well the results extracted from the normal UC in Ref.~\cite{PTM-21}.

As was noted in the Introduction, for the case of open BC, the sign of $b$ at $N = N_c$ is expected to determine which of the two scenarios in Fig.~\ref{fig:RG} for the evolution of the boundary phase diagram with $N$ is realized.  Reference \cite{KM-23} found that, for open BC at large $N$, $b_{\rm OBC}(N) \approx -\frac{4}{3 (2\pi)^2}N$. The present simulations indicate that $b_{\rm OBC}(N)$ is also negative for $N = 2, 3$, which suggests that $b(N)$ remains negative for all $N \ge 2$, i.e. the scenario in Fig.~\ref{fig:RG}, left, is realized. In the language of the microscopic model (\ref{model}) this means that the location of  the special transition $\beta_{s,c} \to \infty$ as $N \to N^-_{c}$, and for $N > N_c$  the ordinary UC is realized for any finite surface coupling $\beta_{s}$.
In Table \ref{tab:b} we compare our results for $b$ in the open BC and defect geometries to the leading large $N$ prediction listed in columns $b^{N}_{\rm OBC}$ and $b^{N}_{\rm def}$. In the defect case, Ref.~\cite{KM-23} found $b_{{\rm def}}(N) \approx -\frac{5}{3 (2\pi)^2} N$, as $N \to \infty$. The two qualitative features of the large $N$ prediction, the sign of $b(N)$ and the increase of $|b(N)|$ with $N$, are manifest in the simulation results.  In fact, the values of $b(N)$ found in the simulation are generally in a reasonable agreement with the leading large $N$ prediction. 

We note that while Ref.~\cite{KM-23} was able to extract the value of $b(N)$ as $N \to \infty$ for both the boundary and defect geometries, a theoretical understanding of $b$ at finite $N$ is still lacking. It is clear that $b$ must be, at least in part, determined by the four point function of the ``tilt" operator for the normal boundary UC (just as the leading coefficient $\alpha$ in the $\beta$-function is determined by the norm of the two-point function of the tilt \cite{PKMGM-21}), however, the precise relation has not been worked out to date. It may be that the coefficients $b(N)$ for the boundary geometry and defect geometry are related (just as the coefficients $\alpha$ are \cite{KM-23}), which could be tested against our MC results. 

In the rest of this section we focus on the open BC geometry. From here on we  assume that for  open BC, $b(N) < 0$ for all $N$. As was pointed out in the introduction, the special transition then becomes perturbatively accessible as $N \to N^-_c$. In this regime, to leading order, the coupling at the special transition is $g_* \approx -\frac{\alpha}{b}$. The scaling dimension of the surface order parameter $\Delta_{\vec{n}}$ and the correlation length exponent $\nu$ then obey \cite{Metlitski-20}
\begin{equation} \Delta_{\vec{n}} \approx \frac{(N-1) g_{*}}{4\pi}, \quad\quad \nu^{-1} \approx \alpha g_*. \label{Deltan} \end{equation}
In addition, we may use Eq.~(\ref{U4g0}) to compute the universal Binder ratio at the special transition:
\begin{equation}  U_4 - 1 \approx \frac{2 (N-1) {\cal S} g^2_*}{(2\pi)^4}. \label{Ucr}\end{equation}
The special transition for $N = 2, 3, 4$ has been numerically studied  in Refs.~\cite{DBN-05}, \cite{PT-20}, \cite{Deng-06}. We may ask whether for these values of $N$ the special transition already lies in the perturbative regime. To assess this, we use the MC results for  $\Delta_{\vec{n}}$, $\nu^{-1}$ and $U_4$ at the special transition listed in Table \ref{tab:sp}, as well as the value of $\alpha$ in Table \ref{tab:alpha},  and  Eqs.~(\ref{Deltan}) and (\ref{Ucr}) to extract the corresponding values of $b$ that we label $b^{\Delta}, b^{\nu}, b^{U}$. We then check the consistency of these values of $b$ among themselves and with the estimate of $b$ in Sec.~\ref{sec:results:beta}, see Table \ref{tab:b}. For $N = 2$, $b^{\nu}$ and $b^{\Delta}$, $b^{U}$ differ considerably among themselves and also disagree with the value $b^{\rm MC}$ extracted in Sec.~\ref{sec:results:beta}. This suggests that for $N = 2$  the special transition is not perturbatively accessible. For $N= 3$, the values $b^{\nu}$, $b^{\Delta}$, $b^U$ and $b^{\rm MC}$ are much closer together, which suggests that in this case leading order perturbation theory describes the special transition to 10-20\% accuracy.  

The fact that the special transition at $N = 3$ can be accessed by perturbation theory has the following consequence. Certain bulk quantum critical points of 2D quantum antiferromagnets map to the 3D classical spacetime $O(3)$ model. However, the effective 1+1D non-linear $\sigma$-model (\ref{nls}) describing the edge of such a quantum system may possess a topological $\theta$ term depending on the value of the microscopic spin and the particular edge termination. In an isolated 1+1D system, such a $\theta$ term plays a crucial role, making an $s=1/2$ spin chain ($\theta = \pi$) gapless compared to the gapped  $s = 1$ spin chain ($\theta = 0$). 
However, MC simulations \cite{ZW-17, DZG-18, WPTW-18, WW-19, WZG-22}  of the  ``dangling edge" of a 2+1D critical quantum magnet appear to show the same critical exponent $\Delta_{\vec{n}}$ for $s = 1/2$ and $s = 1$ systems, which, in fact, approximately agrees with the exponent seen at the special transition in the classical 3D $O(3)$ model \cite{PT-20}.  In the scenario \cite{DZG-18,JXWX-20} where the edge critical behavior in Refs.~\cite{ZW-17, DZG-18, WPTW-18, WW-19, WZG-22} is controlled by the special transition,  Ref.~\cite{Metlitski-20} argued that the insensitivity of the results to the value of the spin ($\theta$) may be caused by the exponential suppression of instantons when the critical coupling $g_*$ of the special transition is small. We caution, however, that unlike for the case of boundary criticality in the 3D classical $O(N)$ model, where in the past several years MC simulations have converged with the theory, the edge critical behavior of quantum magnets remains understood rather poorly. In particular, it is still not fully clear why the IR unstable special UC appears to control a region of the edge parameter space in several microscopic models. In addition, a systematic instanton calculation for the boundary of the 3D $O(3)$ model has not been carried out to date and remains an interesting target for theory.

Returning to the classical 3D $O(N)$ model, we expect perturbation theory to describe the special transition even more accurately for $N=4$. However, in this case the special transition is  more difficult to study with MC than for $N = 3$  due to the smallness of the exponent $\nu^{-1}$ --- a fact one must remember when using the results in Table \ref{tab:sp}. Nevertheless, for $N = 4$ the values $b^{\Delta}$, $b^{\nu}$ and $b^U$ agree quite well.  It would be interesting to repeat the analysis of the special transition for $N=4$ using our ``improved" lattice model to increase the precision of the critical data in Table \ref{tab:sp}. Note that for $N=4$ we have not extracted $b$ from the FSS of RG invariants in the extraordinary-log phase, but $b^{\Delta}$ and $b^U$ agree within 10-20\%  with the leading large $N$ estimate of $b$.

\begin{acknowledgments}
F.P.T. is funded by the Deutsche Forschungsgemeinschaft (DFG, German Research Foundation), Project No. 414456783.
F.P.T. gratefully acknowledges the Gauss Centre for Supercomputing e.V. \cite{gauss} for funding this project by providing computing time through the John von Neumann Institute for Computing (NIC) on the GCS Supercomputer JUWELS at Jülich Supercomputing Centre (JSC).
F.P.T. also gratefully acknowledge computing time granted by the IT Center of RWTH Aachen University and used for the calculations of Appendix \ref{app:beta_O32d}.
A.K. is supported
by the National Science Foundation Graduate Research Fellowship under Grant No.\ 1745302. A.K.
also acknowledges support from the Paul and Daisy Soros Fellowship and the Barry M. Goldwater
Scholarship Foundation. MM is supported in part by the National Science Foundation under Grant No. DMR-1847861. This research was supported in part by Grant No. NSF PHY-2309135 to the Kavli Institute for Theoretical Physics (KITP). We thank the organizers of the focus session ``Recent Progresses in Criticality in the Presence of Boundaries and Defects" at the  DPG Spring Meeting 2024 in Berlin, the  KITP  conference ``Theories, Experiments and Numerics on Gapless Quantum Many-body Systems" and of the workshop ``Defects: from Condensed Matter to Quantum Gravity" at the  Pollica Physics Centre for allowing the authors to interact in person. 
\end{acknowledgments}

\appendix

\section{\texorpdfstring{$U_4$}{U4} in the non-linear \texorpdfstring{$\sigma$}{sigma}-model}
\label{app:U4}
In this appendix we compute the Binder ratio (\ref{U4def}) in the non-linear $\sigma$-model (\ref{nls}) in perturbation theory. As in Sec.~\ref{sec:fss}, writing $\vec{n} = (\vec{\pi}, \sqrt{1- \vec{\pi}^2})$, with $\int d^2 x  \, \vec{\pi} = 0$, we have
\begin{equation}
M_{2p} = \bigg\langle \left(\int d^2 x d^2 y \, \sqrt{1- \vec{\pi}^2(x)}  \sqrt{1 - \vec{\pi}^2(y)} \right)^p  \bigg\rangle.
\end{equation}
Expanding $\sqrt{1- \vec{\pi}^2} = 1 - \frac12 \vec{\pi}^2 - \frac18 (\vec{\pi}^2)^2\ldots$, we obtain
\begin{align}
    M_2 \approx& L^4\left(1 - \langle \vec{\pi}^2\rangle - \frac{1}{4} \langle (\vec{\pi}^2)^2\rangle\right) \nonumber \\
    &+ \frac{1}{4} \int d^2 x d^2 y \langle \vec{\pi}^2(x) \vec{\pi}^2(y)\rangle \\ 
    M_4 \approx& L^8\left(1 - 2 \langle \vec{\pi}^2\rangle - \frac{1}{2} \langle (\vec{\pi}^2)^2\rangle\right) \nonumber \\
    &+ \frac{3}{2} L^4 \int d^2 x d^2 y \langle \vec{\pi}^2(x) \vec{\pi}^2(y)\rangle
\end{align}
Inserting these results in the definition of $U_4$ [Eq.~(\ref{U4def})] we find
\begin{equation}
    \begin{split}
       U_4 - 1 &\approx \frac{1}{L^4} \int d^2x d^2 y \left( \langle \vec{\pi}^2 (x) \vec{\pi}^2(y)\rangle -  \langle \vec{\pi}^2(x) \rangle \langle \vec{\pi}^2(y)\rangle\right )  \\
        &\approx \frac{2 (N-1)}{L^4} \int d^2x d^2 y D^2_{\pi}(x-y) \\
        &= 2 (N-1) g^2 \sum_{\vec{k} \neq 0} \frac{1}{(L^2 \vec{k}^2)^2},
    \end{split}
    \label{U4minus1_partial}
\end{equation}
with $\vec{k} = \frac{2 \pi}{L} (n, m)$ and $n, m \in \mathbb{Z}$. 
The sum in the previous equation can be evaluated as
\begin{equation}
    \begin{split}
    {\cal S} &= \sum_{(n, m) \neq (0,0)} \frac{1}{(n^2 + m^2)^2} = 4 \sum_{n = 1}^{\infty} \frac{1}{n^4} + 4 \sum_{n = 1}^{\infty} \sum_{m=1}^{\infty} \frac{1}{(n^2 + m^2)^2} \\
     &= 4 \zeta (4) + 4\sum_{m=1}^{\infty} \left(\frac{\pi}{4 m^3} \coth \pi m + \frac{\pi^2}{4 m^2 \sinh^2(\pi m)} - \frac{1}{2 m^4}\right) \\
     &\approx 6.02681.      
    \end{split}
\label{sumU4}
\end{equation}
Finally, inserting Eq.~(\ref{sumU4}) in Eq.~(\ref{U4minus1_partial}) we arrive at Eq.~(\ref{U4g0}).

\section{\texorpdfstring{$\beta-$}{Beta-}function determination for \texorpdfstring{$s=3/2$}{s=3/2}}
\label{app:beta_s1.5}
\begin{figure*}[t]
    \centering
    \includegraphics[width=\linewidth]{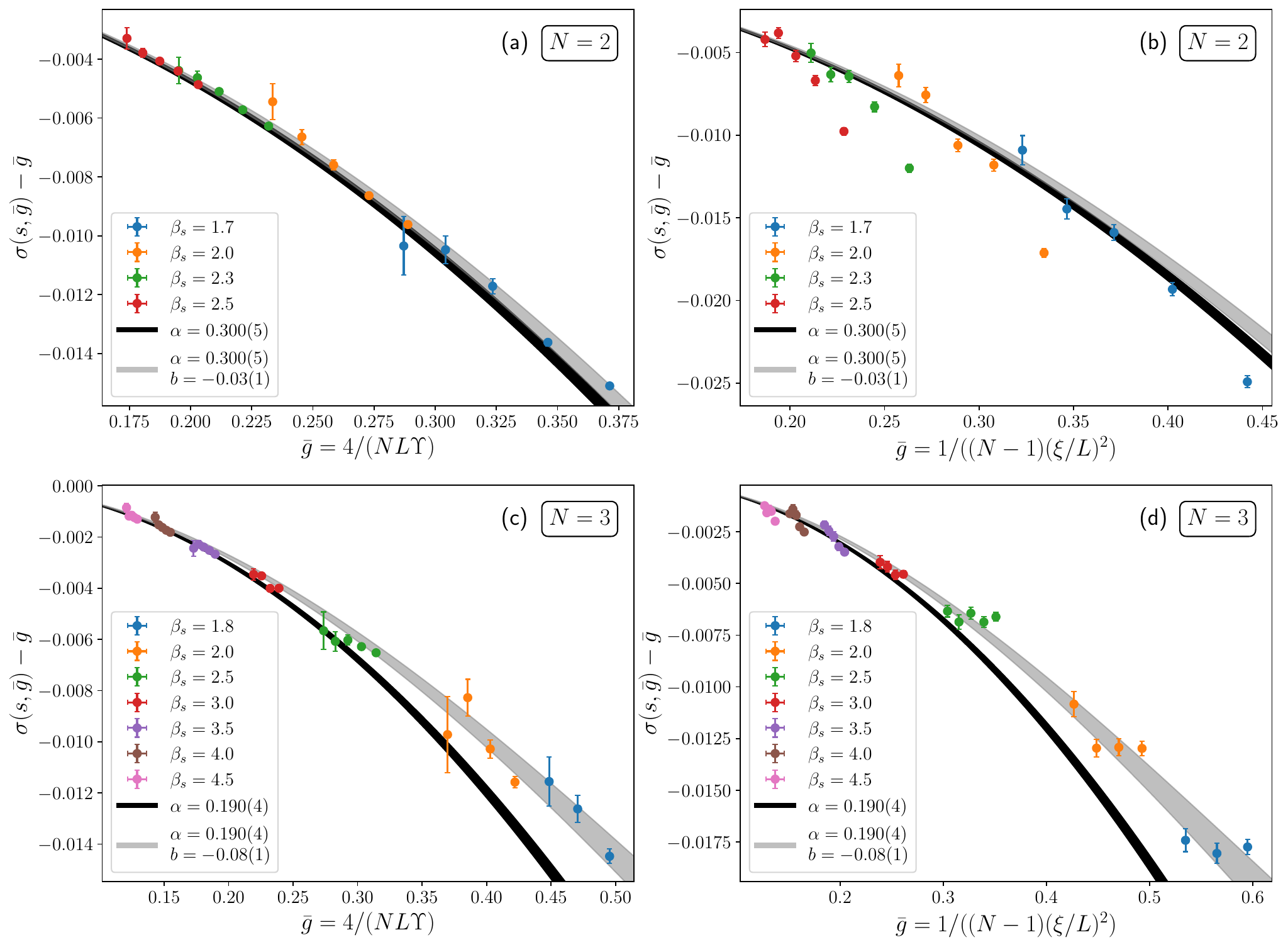}
    \caption{Same as Fig.~\ref{fig:RGflow_open_s2} but with scale factor $s = 3/2$.
    }
    \label{fig:RGflow_open_s1.5}
\end{figure*}
\begin{figure*}[t]
    \centering
    \includegraphics[width=\linewidth]{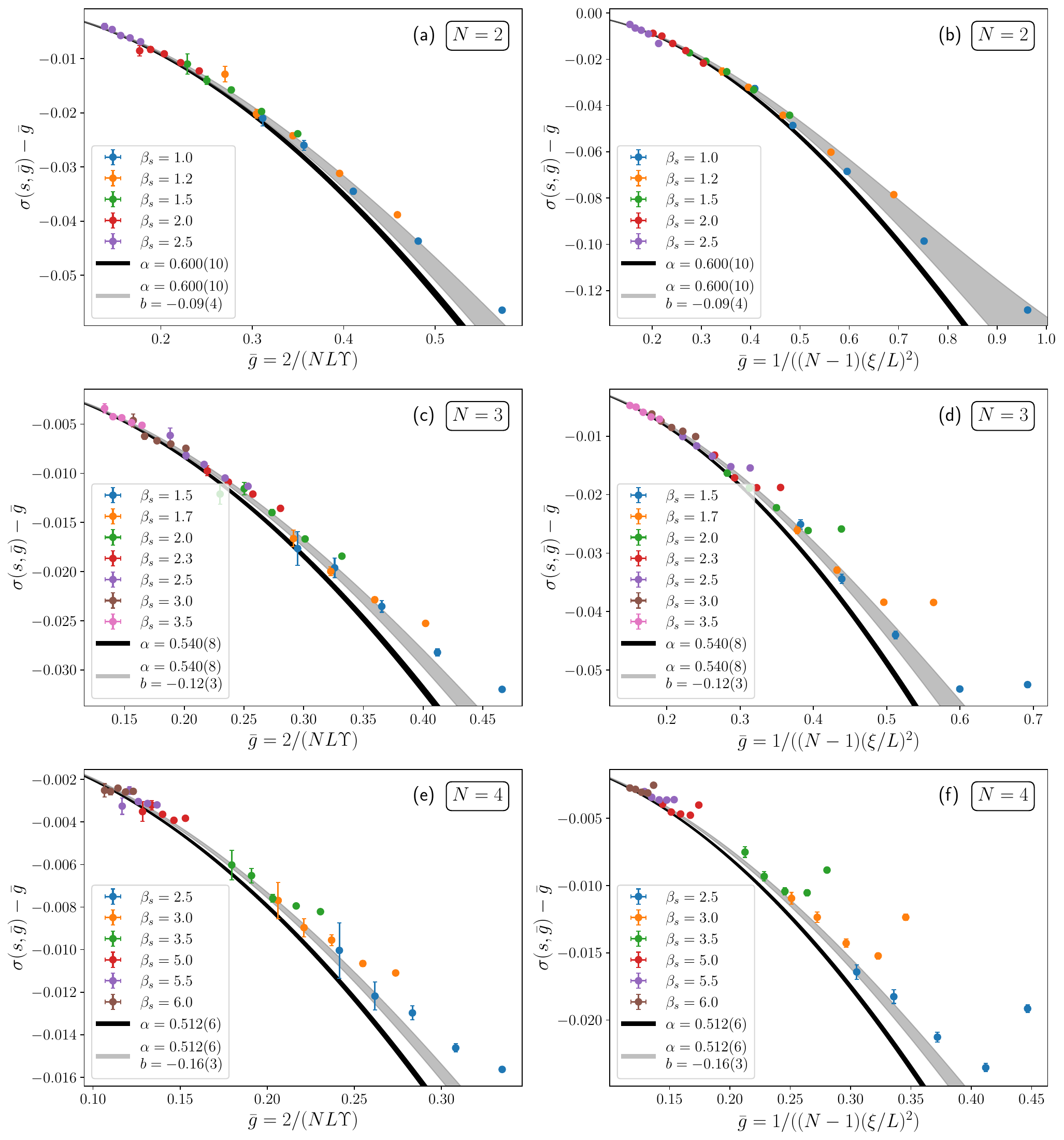}
    \caption{Same as Fig.~\ref{fig:RGflow_open_s2} but with scale factor $s = 3/2$ and for the plane-defect geometry.
    }
    \label{fig:RGflow_plane_s1.5}
\end{figure*}

In this appendix we repeat the procedure outlined in Sec.~\ref{sec:results:beta} to compute the $\beta-$function, setting the scaling parameter $s=3/2$.
In Figs.~\ref{fig:RGflow_open_s1.5} and \ref{fig:RGflow_plane_s1.5} we show the step scaling function $\sigma(s=3/2, \bar{g})$ for open BCs and for the plane-defect geometry.
Compared to the case $s=2$ studied in Sec.~\ref{sec:results:beta}, due to the smaller step $s$ the $y-$scale on Figs.~\ref{fig:RGflow_open_s1.5} and \ref{fig:RGflow_plane_s1.5} is smaller.
Furthermore, with the available lattice sizes, less points are present.
Besides this, the qualitative features of $\sigma(s=3/2, \bar{g})$ are similar to the $s=2$ case.
In particular, we observe significantly more finite-size corrections when $\bar{g}$ is defined in terms of $\xi/L$.
As in the case $s=2$, in Figs.~\ref{fig:RGflow_open_s1.5} and \ref{fig:RGflow_plane_s1.5} we plot $\sigma(s=3/2,\bar{g})$ in Eq.~(\ref{sigma}) truncated to $O(\bar{g}^3)$ with  value of $\alpha$ given in Table \ref{tab:alpha} and $b$ set either to zero (black range) or it's value $b^{\rm MC}$ in Table \ref{tab:b} (gray range).

We have repeated the quantitative analysis by fitting $\sigma(s=3/2,\bar{g})$ to the right-hand side of Eq.~(\ref{sigma}).
Results are reported in Tables \ref{tab:fits_RGflow_open_s1.5} and \ref{tab:fits_RGflow_plane_s1.5}.
We observe in general a good agreement of the fitted $\alpha$ with the expected values in Table \ref{tab:alpha}.
Overall, due to the more ``compressed" scale in $\sigma(s=3/2, \bar{g})$ and less data points, the fitted coefficients are less precise than the $s=2$ case.

\begin{table*}
    \centering
    \caption{Fits of $\sigma(s=3/2, \bar{g})$ for open BCs, and for $N=2,3$ to the right-hand side of Eq.~(\ref{sigma}).
    For the renormalized coupling constant $\bar{g}$  defined in Eq.~(\ref{gLY_open}) or in Eq.~(\ref{gxil}),
    we report the resulting coefficients $\alpha$ and $b$ as a function of maximum value $\bar{g}_{\rm max}$ of $\bar{g}$ and of the minimum lattice size $L_{\rm min}$ employed in the fit.
    }
    \begin{ruledtabular}
    \begin{tabular}{llaabllaab}
       \multicolumn{5}{c}{$\bar{g} = 4 / (N L\Upsilon)$}  & \multicolumn{5}{c}{$\bar{g} = 1 / [(N-1)(\xi/L)^2]$} \\
       \cline{1-5} \cline{6-10}
       \multicolumn{1}{l}{\rule{0pt}{2.5ex}$\bar{g}_{\rm max}$} & \multicolumn{1}{l}{$L_{\rm min}$} & \multicolumn{1}{c}{$\alpha$}  & \multicolumn{1}{c}{$b$} & \multicolumn{1}{c}{$\chi^2/\text{d.o.f.}$} &
       \multicolumn{1}{l}{$\bar{g}_{\rm max}$} & \multicolumn{1}{l}{$L_{\rm min}$} & \multicolumn{1}{c}{$\alpha$}  & \multicolumn{1}{c}{$b$} & \multicolumn{1}{c}{$\chi^2/\text{d.o.f.}$}\\
       \hline
       \multicolumn{10}{c}{\rule{0pt}{2.5ex}$N=2$} \\
       \hline\rule{0pt}{2.5ex}
       &   8  &   0.3011(33)  &   -0.023(12)  & 0.9    &        &   8  &    0.382(18)  &   -0.049(58)  & 27.5 \\
  0.35 &  16  &   0.2935(50)  &   -0.002(18)  & 0.7    &   0.35 &  16  &    0.336(24)  &   -0.074(82)  & 4.6 \\
       &  32  &   0.292(12)   &   -0.014(47)  & 0.4    &        &  32  &    0.280(28)  &    0.064(97)  & 2.8 \\[1em]

       &   8  &   0.3000(40)  &   -0.018(16)  & 1.1    &        &   8  &    0.382(36)  &    -0.04(13)  & 29.7 \\
   0.3 &  16  &   0.2857(91)  &    0.032(38)  & 0.7    &    0.3 &  16  &    0.340(39)  &    -0.09(15)  & 5.6 \\
       &  32  &    0.295(22)  &   -0.029(94)  & 0.5    &        &  32  &    0.269(42)  &     0.11(16)  & 3.1 \\[1em]

       &   8  &    0.300(11)  &   -0.018(47)  & 1.3    &        &   8  &    0.057(75)  &     1.35(33)  & 21.1 \\
  0.25 &  16  &    0.287(21)  &    0.028(97)  & 0.8    &   0.25 &  16  &    0.129(77)  &     0.86(33)  & 3.2 \\
       &  32  &    0.329(35)  &    -0.19(16)  & 0.3    &        &  32  &     0.16(12)  &     0.62(54)  & 1.3 \\[1em]

       &   8  &     0.27(11)  &     0.13(53)  & 0.1    &        &   8  &    -0.13(42)  &     2.1(2.2)  & 3.5 \\
   0.2 &  16  &     0.27(11)  &     0.13(53)  & 0.1    &   0.21 &  16  &    -0.13(42)  &     2.1(2.2)  & 3.5 \\
       &  32  &     0.23(24)  &     0.3(1.2)  & 0.2    &        &  32  &    -0.13(42)  &     2.1(2.2)  & 3.5 \\
       \multicolumn{10}{c}{\rule{0pt}{2.5ex}$N=3$} \\
       \hline\rule{0pt}{2.5ex}
       &   8  &   0.2099(20)  &  -0.1286(89)  & 1.3    &        &   8  &   0.2588(76)  &   -0.312(25)  & 3.0 \\
   0.4 &  16  &   0.2014(39)  &   -0.086(18)  & 1.1    &    0.4 &  16  &   0.2229(99)  &   -0.187(34)  & 1.6 \\
       &  32  &   0.1950(72)  &   -0.073(32)  & 1.0    &        &  32  &    0.198(12)  &   -0.103(44)  & 1.3 \\[1em]

       &   8  &   0.2083(29)  &   -0.119(15)  & 1.3    &        &   8  &    0.262(13)  &   -0.332(59)  & 3.2 \\
   0.3 &  16  &   0.1953(50)  &   -0.050(26)  & 0.9    &    0.3 &  16  &    0.215(17)  &   -0.152(81)  & 1.7 \\
       &  32  &   0.1868(82)  &   -0.028(39)  & 0.7    &        &  32  &    0.202(22)  &    -0.13(11)  & 1.2 \\[1em]

       &   8  &   0.2079(52)  &   -0.116(32)  & 0.9    &        &   8  &    0.288(29)  &    -0.51(16)  & 3.9 \\
   0.2 &  16  &    0.202(10)  &   -0.096(64)  & 0.8    &    0.2 &  16  &    0.226(32)  &    -0.22(18)  & 2.1 \\
       &  32  &    0.183(22)  &   0.005(137)  & 0.7    &        &  32  &    0.251(39)  &    -0.43(23)  & 1.2 \\
    \end{tabular}
    \end{ruledtabular}
    \label{tab:fits_RGflow_open_s1.5}
\end{table*}

\begin{table*}
    \centering
    \caption{Same as Table \ref{tab:fits_RGflow_open_s1.5} for the plane-defect geometry.
    }
    \begin{ruledtabular}
    \begin{tabular}{llaabllaab}
       \multicolumn{5}{c}{$\bar{g} = 4 / (N L\Upsilon)$}  & \multicolumn{5}{c}{$\bar{g} = 1 / [(N-1)(\xi/L)^2]$} \\
       \cline{1-5} \cline{6-10}
       \multicolumn{1}{l}{\rule{0pt}{2.5ex}$\bar{g}_{\rm max}$} & \multicolumn{1}{l}{$L_{\rm min}$} & \multicolumn{1}{c}{$\alpha$}  & \multicolumn{1}{c}{$b$} & \multicolumn{1}{c}{$\chi^2/\text{d.o.f.}$} &
       \multicolumn{1}{l}{$\bar{g}_{\rm max}$} & \multicolumn{1}{l}{$L_{\rm min}$} & \multicolumn{1}{c}{$\alpha$}  & \multicolumn{1}{c}{$b$} & \multicolumn{1}{c}{$\chi^2/\text{d.o.f.}$}\\
       \hline
       \multicolumn{10}{c}{\rule{0pt}{2.5ex}$N=2$} \\
       \hline\rule{0pt}{2.5ex}
       &   8  &   0.5894(37)  &   -0.149(12)  & 3.7    &        &   8  &   0.6398(72)  &   -0.138(12)  & 10.0 \\
   0.4 &  16  &   0.5977(70)  &   -0.130(21)  & 1.2    &    0.6 &  16  &   0.6026(81)  &   -0.080(13)  & 2.5 \\
       &  32  &   0.620(17)   &   -0.188(56)  & 1.3    &        &  32  &    0.588(10)  &   -0.047(17)  & 1.4 \\[1em]

       &   8  &   0.5980(86)  &   -0.187(37)  & 2.8    &        &   8  &    0.711(13)  &   -0.327(32)  & 8.6 \\
   0.3 &  16  &    0.614(17)  &   -0.201(78)  & 1.5    &   0.45 &  16  &    0.648(14)  &   -0.204(35)  & 1.3 \\
       &  32  &    0.659(27)  &    -0.38(11)  & 1.3    &        &  32  &    0.624(21)  &   -0.160(57)  & 1.1 \\[1em]

       &   8  &    0.758(58)  &    -1.02(29)  & 2.2    &        &   8  &    0.719(35)  &    -0.37(13)  & 13.8 \\
   0.2 &  16  &     0.64(13)  &    -0.38(70)  & 2.5    &    0.3 &  16  &    0.616(37)  &    -0.08(14)  & 2.1 \\
       &  32  &     0.44(15)  &     0.96(90)  & 1.3    &        &  32  &    0.590(53)  &    -0.02(21)  & 1.7 \\
       \multicolumn{10}{c}{\rule{0pt}{2.5ex}$N=3$} \\
       \hline\rule{0pt}{2.5ex}
       &   8  &   0.5229(39)  &   -0.198(14)  & 9.0    &        &   8  &   0.5408(45)  &  -0.2286(78)  & 34.2 \\
   0.4 &  16  &   0.5405(71)  &   -0.171(24)  & 2.0    &    0.6 &  16  &   0.5601(55)  &  -0.1983(94)  & 3.3 \\
       &  32  &    0.556(14)  &   -0.183(47)  & 1.5    &        &  32  &   0.5557(93)  &   -0.154(20)  & 1.1 \\[1em]

       &   8  &   0.5379(56)  &   -0.266(22)  & 7.2    &        &   8  &   0.5814(75)  &   -0.341(18)  & 19.9 \\
   0.3 &  16  &    0.558(12)  &   -0.250(48)  & 2.0    &   0.45 &  16  &   0.5522(96)  &   -0.174(24)  & 2.5 \\
       &  32  &    0.565(21)  &   -0.226(86)  & 1.5    &        &  32  &   0.558(11)   &   -0.159(28)  & 1.1 \\[1em]

       &   8  &    0.442(33)  &     0.28(19)  & 8.2    &        &   8  &    0.524(19)  &   -0.095(69)  & 4.0 \\
   0.2 &  16  &    0.513(35)  &  -0.006(190)  & 2.7    &    0.3 &  16  &    0.539(20)  &   -0.111(73)  & 2.1 \\
       &  32  &    0.392(77)  &     0.80(44)  & 1.7    &        &  32  &    0.517(24)  &    0.009(89)  & 0.7 \\
       \multicolumn{10}{c}{\rule{0pt}{2.5ex}$N=4$} \\
       \hline\rule{0pt}{2.5ex}
       &   8  &   0.4762(34)  &   -0.280(16)  & 11.4    &        &   8  &   0.5163(74)  &   -0.492(24)  & 29.9 \\
   0.3 &  16  &   0.5035(68)  &   -0.268(33)  & 1.4    &   0.35 &  16  &   0.5114(92)  &   -0.314(32)  & 3.8 \\
       &  32  &    0.514(12)  &   -0.267(58)  & 1.2    &        &  32  &    0.509(12)  &   -0.229(44)  & 0.9 \\[1em]

       &   8  &   0.4746(47)  &   -0.271(27)  & 11.2   &        &   8  &    0.442(16)  &   -0.079(90)  & 12.2 \\
  0.25 &  16  &   0.4987(90)  &   -0.236(52)  & 1.5    &   0.25 &  16  &    0.489(17)  &   -0.173(92)  & 1.7 \\
       &  32  &   0.504(16)   &   -0.202(86)  & 1.2    &        &  32  &    0.514(19)  &   -0.255(96)  & 0.8 \\[1em]

       &   8  &    0.508(15)  &    -0.51(10)  & 11.2    &        &   8  &    0.736(95)  &    -2.03(64)  & 6.9 \\
   0.2 &  16  &    0.486(26)  &    -0.14(18)  & 1.4    &   0.15 &  16  &    0.706(94)  &    -1.70(64)  & 0.7 \\
       &  32  &    0.508(38)  &    -0.24(27)  & 1.5    &        &  32  &     0.61(12)  &    -0.99(84)  & 0.2 \\
    \end{tabular}
    \end{ruledtabular}
    \label{tab:fits_RGflow_plane_s1.5}
\end{table*}

\section{O(\texorpdfstring{$3$}{3}) model in \texorpdfstring{$d=2$}{d=2}}
\label{app:beta_O32d}

\begin{figure*}
    \centering
    \includegraphics[width=\linewidth]{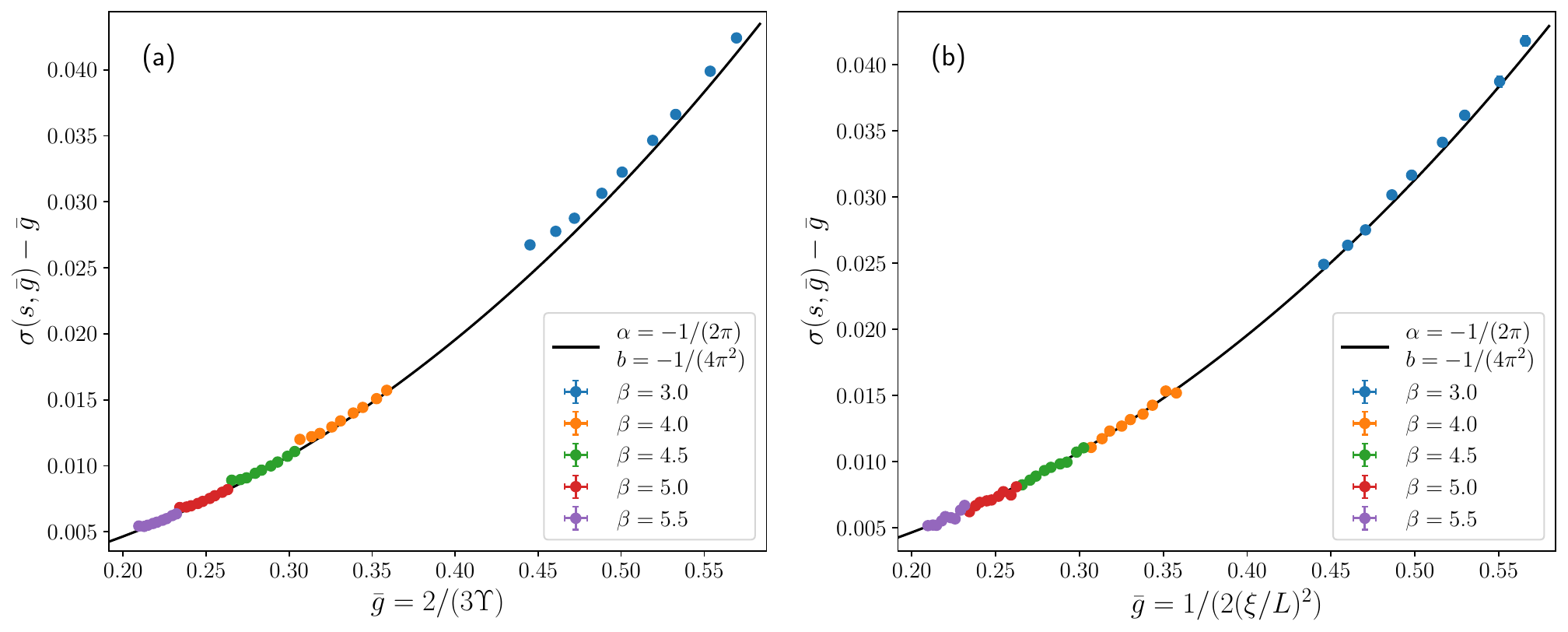}
    \caption{Step scaling function $\sigma(s, \bar{g})$ subtracted of the trivial linear term [See Eq.~(\ref{sigma})] for a 2D O(3) model, $s=2$ and at low temperatures.
    (a): $\sigma(s, \bar{g})$ obtained with $\bar{g}$ defined in Eq.~(\ref{gY_2d});
    (b): $\sigma(s, \bar{g})$ obtained with $\bar{g}$ defined in Eq.~(\ref{gxil}).
    For each value of $\beta_s$, points with smaller $\bar{g}$ correspond to the smallest lattices.
    The solid line corresponds to $\sigma(s, \bar{g})$ truncated to the third order [see Eq.~(\ref{sigma})] and computed using the indicated known values of $\alpha$ and $b$ given in Eq.~(\ref{alpha_b_O32d}).}
    \label{fig:RGflow_O32d}
\end{figure*}

As a test of the ``step-scaling-function" method used to extract the $\beta$-function in Sec.~\ref{sec:results:beta}, we study here the O($3$) model in two dimensions. 
In fact, this method was originally introduced in  the context for the 2D O($3$) model in Ref.~\cite{LWW-91} and later used in Refs.~\cite{Kim-93,CEPS-95,MW-01} . Here we would like to directly compare our data on the 3D model with a boundary/defect and the 2D model, utilizing the same RG invariants (spin-stiffness and correlation length ratio) and with comparable system sizes/invariant range.

The  reduced Hamiltonian of the 2D O($3$) model that we study is
\begin{equation}
    {\cal H} = -\beta\sum_{\< i\ j\>}\vec{S}_i\cdot\vec{S}_j,
    \label{model2d}
\end{equation}
where $\vec{S}$ is a three-components unit vector $|\vec{S}|=1$ and the sum extends over nearest-neighbors sites on a two-dimensional lattice.
The reduced Hamiltonian (\ref{model2d}) corresponds to the limit $\lambda\rightarrow\infty$ of Eq.~(\ref{model}) for $N=3$ and in the absence of boundaries.
The method to compute the $\beta-$functions is identical to that of Sec.~\ref{sec:results:beta}, with the only distinction that, due the different dimensionality, the helicity modulus $\Upsilon$, rather than the combination $L\Upsilon$, is in this case a RG-invariant quantity.
Its relation to the renormalized coupling $\bar{g}$ is
\begin{equation}
    \bar{g} = \frac{2}{N\Upsilon} = \frac{2}{3\Upsilon}.
    \label{gY_2d}
\end{equation}
The first coefficients of the $\beta-$function (\ref{beta}) of the 2D O($3$) model are known \cite{Zinn-book}:
\begin{equation}
    \alpha = -\frac{1}{2\pi} \simeq -0.1592,\qquad
    b = -\frac{1}{4\pi^2} \simeq -0.0253.
\label{alpha_b_O32d}
\end{equation}
Accordingly, the RG-flow does not have an IR stable fixed point and the coupling constant flows to $g=\infty$, i.e., to a disordered state.
We have sampled the model (\ref{model2d}) for a few values of the coupling constant $\beta$ and lattice sizes $8\le L \le 256$.
In Fig.~\ref{fig:RGflow_O32d} we show the resulting step scaling function (\ref{sigmadef}), computed with scaling factor $s=2$. 
A comparison with the scaling function (\ref{sigma}) computed using Eq.~(\ref{alpha_b_O32d}) and up to the third order reveals a rather good agreement with MC data.

\begin{table*}
    \centering
    \caption{Fits of $\sigma(s=2, \bar{g})$ for the O($3$) model in two dimensions to the right-hand side of Eq.~(\ref{sigma}).
    For the renormalized coupling constant $\bar{g}$  defined in Eq.~(\ref{gY_2d}) or in Eq.~(\ref{gxil}),
    we report the resulting coefficients $\alpha$ and $b$ as a function of maximum value $\bar{g}_{\rm max}$ of $\bar{g}$ and of the minimum lattice size $L_{\rm min}$ employed in the fit.
    }
    \begin{ruledtabular}
    \begin{tabular}{llaabllaab}
       \multicolumn{5}{c}{$\bar{g} = 2 / (3\Upsilon)$}  & \multicolumn{5}{c}{$\bar{g} = 1 / [2(\xi/L)^2]$} \\
       \cline{1-5} \cline{6-10}
       \multicolumn{1}{l}{\rule{0pt}{2.5ex}$\bar{g}_{\rm max}$} & \multicolumn{1}{l}{$L_{\rm min}$} & \multicolumn{1}{c}{$\alpha$}  & \multicolumn{1}{c}{$b$} & \multicolumn{1}{c}{$\chi^2/\text{d.o.f.}$} &
       \multicolumn{1}{l}{$\bar{g}_{\rm max}$} & \multicolumn{1}{l}{$L_{\rm min}$} & \multicolumn{1}{c}{$\alpha$}  & \multicolumn{1}{c}{$b$} & \multicolumn{1}{c}{$\chi^2/\text{d.o.f.}$}\\
       \hline\\
       &   8  & -0.15834(14)  & -0.04512(51)  & 189.7    &        &   8  & -0.15535(67)  &  -0.0382(22)  & 1.0 \\
       &  12  & -0.15741(18)  & -0.04243(66)  & 29.6    &        &  12  & -0.15590(87)  &  -0.0369(29)  & 1.0 \\
   0.5 &  16  & -0.15793(24)  & -0.03793(91)  & 9.2    &    0.5 &  16  &  -0.1561(11)  &  -0.0365(36)  & 1.1 \\
       &  24  & -0.15755(34)  &  -0.0375(12)  & 2.9    &        &  24  &  -0.1548(16)  &  -0.0407(54)  & 1.1 \\
       &  32  & -0.15780(51)  &  -0.0355(19)  & 1.5    &        &  32  &  -0.1555(21)  &  -0.0379(72)  & 1.3 \\
       &  48  & -0.15798(74)  &  -0.0336(27)  & 0.5    &        &  48  &  -0.1563(43)  &   -0.034(15)  & 1.5 \\[1em]

       &   8  & -0.16143(21)  & -0.03264(85)  & 165.5    &        &   8  &  -0.1553(14)  &  -0.0385(52)  & 1.0 \\
       &  12  & -0.15903(27)  &  -0.0360(11)  & 24.3    &        &  12  &  -0.1555(17)  &  -0.0384(66)  & 1.0 \\
  0.35 &  16  & -0.15864(34)  &  -0.0352(13)  & 7.7    &   0.35 &  16  &  -0.1553(21)  &  -0.0398(79)  & 1.1 \\
       &  24  & -0.15776(47)  &  -0.0368(18)  & 2.4    &        &  24  &  -0.1579(29)  &   -0.029(11)  & 1.0 \\
       &  32  & -0.15719(60)  &  -0.0381(23)  & 1.3    &        &  32  &  -0.1583(36)  &   -0.028(13)  & 1.2 \\
       &  48  & -0.15746(91)  &  -0.0358(35)  & 0.5    &        &  48  &  -0.1577(55)  &   -0.028(20)  & 1.4 \\[1em]

       &   8  &  -0.1581(10)  &  -0.0495(48)  & 176.4    &        &   8  &  -0.1586(84)  &   -0.024(39)  & 1.5 \\
       &  12  &  -0.1547(14)  &  -0.0574(66)  & 24.0    &        &  12  &   -0.141(11)  &   -0.105(53)  & 1.2 \\
  0.25 &  16  &  -0.1557(19)  &  -0.0498(87)  & 8.6    &   0.25 &  16  &   -0.140(15)  &   -0.110(70)  & 1.5 \\
       &  24  &  -0.1573(31)  &   -0.040(14)  & 2.7    &        &  24  &   -0.165(24)  &   0.003(111)  & 1.6 \\
       &  32  &  -0.1548(45)  &   -0.050(21)  & 0.8    &        &  32  &   -0.181(35)  &     0.07(16)  & 2.2 \\
    \end{tabular}
    \end{ruledtabular}
    \label{tab:fits_RGflow_O32d}
\end{table*}
For a more quantitative test, we have fitted the step scaling function to Eq.~(\ref{sigma}).
In Table \ref{tab:fits_RGflow_O32d} we report the fit results.
Fits of the helicity modulus show overall a relatively large $\chi^2/\text{d.o.f.}$, and a residual difference between the fitted parameters and the expected values of Eq.~(\ref{alpha_b_O32d}).
We also observe a small but significant trend in the fitted values: on increasing the smallest lattice size used in the fits, $\alpha$ appears to increase and $b$ to decrease.
Fits of the the correlations length show a better agreement with the expected values: with $\bar{g}_\text{max} = 0.35$, results for $L_{\text{min}}\ge 24$ are in perfect agreement with Eq.~(\ref{alpha_b_O32d}). It should be noted that overall the precision is smaller than in the fits of $\Upsilon$. Note that the expected coefficient $b$ for the 2D O($3$) model is rather small, close in magnitude to our estimate of $b$ for the boundary of the 3D O($2$) model in Sec.~\ref{sec:results:beta}. The performance of the method for the 2D O($3$) model suggests that it reliably gives the sign of $b$ and a reasonable estimate of its  magnitude.

\bibliography{short}

\end{document}